\documentclass[article, prd,nofootinbib,floats,superscriptaddress,eqsecnum,tightenlines,11pt]{revtex4}

\usepackage{hyperref}
\usepackage{graphicx}
\usepackage{amsmath,amssymb,amsfonts,amsthm,latexsym,stmaryrd,bbm}
\usepackage{marginnote}
\usepackage[normalem]{ulem}
\usepackage{color}
\usepackage{braket}

\def\be{\begin{equation}}
\def\ee{\end{equation}}
\def\bea{\begin{eqnarray}}
\def\eea{\end{eqnarray}}
\def\bas{\begin{subequations}\begin{eqnarray}}
\def\eas{\end{eqnarray}\end{subequations}}
\def\nn{\nonumber}

\newcommand{\cG}{{\mathcal G}}

\newcommand{\mat} [2] {\left ( \begin{array}{#1}#2\end{array} \right ) }

\begin{document}

\title{A new look at scalar perturbations in loop quantum cosmology: \\ (un)deformed algebra approach using self dual variables}

\author{Jibril Ben Achour}\email{jibrilbenachour@gmail.com}
\affiliation{Center for Field Theory and Particle Physics, Fudan University, Shanghai, China}

\author{Suddhasattwa Brahma}\email{suddhasattwa.brahma@gmail.com}
\affiliation{Center for Field Theory and Particle Physics, Fudan University, Shanghai, China}

\author{Julien Grain}\email{julien.grain@ias.u-psud.fr}
\affiliation{CNRS, Institut d'Astrophysique Spatiale, UMR8617, Orsay, F-91405, France}
\affiliation{Universit\'e Paris Sud, Orsay, F-91405, France}

\author{Antonino Marcian\`o}\email{marciano@fudan.edu.cn}
\affiliation{Center for Field Theory and Particle Physics, Fudan University, Shanghai, China}

%\date{\today}

\begin{abstract}
Scalar cosmological perturbations in loop quantum cosmology (LQC) is revisited in a covariant manner, using self dual Ashtekar variables. For real-valued Ashtekar-Barbero variables, this `deformed algebra' approach has been shown to implement holonomy corrections from loop quantum gravity (LQG) in a consistent manner, albeit deforming the algebra of modified constraints in the process. This deformation has serious conceptual ramifications, not the least of them being an effective `signature-change' in the deep quantum regime. In this paper, we show that working with self dual variables lead to an \textit{undeformed} algebra of hypersurface deformations, even after including holonomy corrections in the effective constraints. As a necessary consequence, the diffeomorphism constraint picks up non-perturbative quantum corrections thus hinting at a modification of the underlying space-time structure, a novel ingredient compared to the usual treatment of (spatial) diffeomorphisms in LQG. This work extends a similar result obtained in the context of spherically symmetric gravity coupled to a scalar field, suggesting that self dual variables could be better suited than their real counterparts to treat inhomogeneous LQG models.
\end{abstract}

\maketitle

\section{Introduction}
The loop quantization of homogenous cosmological spacetimes has revealed a new paradigm to study the dynamics of the deep Planckian regime of the universe, through a regular FLRW quantum geometry. The LQC framework since then has been extended to accommodate a perturbed FLRW geometry. The treatment of cosmological perturbations in this non-singular quantum cosmology allows one to extend the inflationary scenario to the very early universe and, thereby, compute the power spectrum of curvature perturbations taking into account the impact of the non-perturbative quantum corrections due to loop quantization. This new paradigm for treating cosmological perturbations on a loop-quantized background provided a rare direct contact between a proposal for a fundamental quantum gravity theory and observational data.

This extension of LQC to the perturbed FLRW spacetime was initiated using two principal strategies: the `dressed' metric approach and the `deformed algebra' approach (We refer to e.g. \cite{Merce} for another perspective on LQC perturbations). The interested reader can refer to \cite{LQCComparison} for a pedagogical overview of these two approaches, as well to \cite{LQCComparison2} for a comparison between the two. In this paper we take the perspective that quantum corrections need to be implemented in a covariant manner, as in the deformed algebra approach.

\subsection{Why is general covariance important for quantum gravity?}
Since our guiding principle throughout is going to be the requirement of (some notion of) general covariance at the effective level, it deserves some justification. Indeed, general covariance has turned out to be a strong constraint for several models of loop quantum gravity in recent times (see, for instance \cite{CovSpher, CovGowdy, CovGen}). In Lagrangian formulations, general covariance in the classical action is assured as long as one properly contracts the Lorentz indices of tensors. However, one has to then check that the quantization procedure does not generate anomalies which violate covariance. However, in the canonical formulation covariance is not immediately manifest and are replaced instead by the invariance under gauge transformations generated by the Hamiltonian constraint $H[N]$ and the diffeomorphism constraints $D[N^a]$. The $N$ and $N^a$ are a smearing (lapse) function and a smearing (shift) vector for these constraints, which classically satisfy the algebra \cite{Dirac}
\bea\label{HDA}
\left\{D\left[N^a_1\right]\,,\,D\left[N^a_2\right]\right\} &=& D\left[\mathcal{L}_{N_1} N^a_2\right]\,,\\
\left\{H\left[N\right]\,,\,D\left[M^a\right]\right\} &=& - H\left[\mathcal{L}_{M} N\right]\,,\\
\left\{H\left[N_1\right]\,,\,H\left[N_2\right]\right\} &=& D\left[q^{ab}\,\left(N_1\nabla_b N_2 - N_2\nabla_b N_1\right)\right]\,.
\eea
The gauge transformations generated by these constraints, on the space of solutions to these constraints, are equivalent to space-time Lie derivatives which implement general coordinate transformations. Thus general covariance in canonical theories is implemented in a much more subtle way than for Lagrangian theories \cite{BojowaldBook}. It is important to remember that covariance, defined in this manner, is a condition that is not restricted by solutions of Einstein's equations as they are an off-shell statement (and thus are not restricted to the constraint surface). This is the same as in the Lagrangian formalism, where we can construct covariant actions from metrics which are not necessarily solutions of general relativity (GR).

However, it is important that quantization of a canonical theory of gravity does not lead to anomalies which break general covariance. The conditions that require a quantization is covariant has been stated in \cite{CovSpher}. They basically state that the algebra of quantum constraints remain a first-class system and that they have a well-defined classical limit, in which one recovers the algebra of hypersurface deformations (\ref{HDA}). In LQG, one requires that connections are replaced by holonomies (or parallel transport of connections) since those are variables well defined on the LQG Hilbert space. Such `polymerization' can lead to several interesting phenomenon such as `signature-change', as has been suggested in theories involving cosmological perturbations of LQC (for instance, see \cite{Bojowald:2015gra}) and Schwarzschild black hole models \cite{CovSpher}. The main result involves deformation of the structure function appearing in the algebra (\ref{HDA}) when holonomy modifications from LQG are introduced in the Hamiltonian constraint. (Hence this is sometimes called the `deformed algebra' approach in LQG since this typically results in deformation of the structure functions in the algebra when working with real-valued variables.) Indeed, it is a nontrivial result to find that the `effective' algebra of constraints remain first class and thus eliminates the same number of spurious degrees of freedom and does not violate gauge invariance. In the formalism involving real variables of LQG, the algebra of constraints remain anomaly free but with such a deformation of the structure function that has the same sign as in Euclidean gravity in higher curvature regimes, thus giving us the notion of `signature change' \cite{Bojowald:2012ux,WDWSS,CovCGHS}.

Instead of requiring that the full algebra of the total constraints (that is, including both matter and gravitational contributions) remain anomaly free, sometimes a background treatment is preferred. In such a case one quantizes the background gravitational system and then puts matter on top of the `quantized' background \cite{QFonQG1, Agullo2, GP1, GP2}. (A specific example of this is the dressed metric approach \cite{Agullo1,Agullo2,Agullo3}.) Even if the quantization of the background has been performed in a covariant manner, and then one checks that local energy-momentum conservation law $\nabla_a T^{ab} = 0$ is satisfied \cite{Casey}, it implies that both the matter and gravitational parts of the constraint algebra forms two first-class systems. Recall, that such a treatment \textit{does} require several restrictions and yet is not as stringent as a background-independent approach that we take here. In our approach, we require that the algebra of the total constraints remain closed, i.e. both the gravitational and matter parts of the constraints not only remain closed but also have \textit{matching} forms of covariance (for more details, see \cite{CovSpher}).

\subsection{Cosmological perturbations with self dual Ashtekar variables}
An overview of the results obtained within the deformed algebra formulation using real-valued Ashtekar-Barbero variables is given in the next section. The crucial outcomes of this approach, relevant for us, are the following. Firstly, the quantum-corrected effective constraints form a closed, yet deformed, algebra of constraints in the presence of holonomy modifications. Interestingly, this deformation enters explicitly in the equation of motion for the perturbations and thus has an important impact on their primordial power spectrum derived in this approach. Moreover, the generator of spatial diffeomorphisms remain unmodified by quantum corrections within this formulation.

In this paper, we revisit scalar cosmological perturbations in the deformed algebra approach with self dual Ashtekar variables. Our work is motivated by a previous result obtained in the case of spherical symmetry: Using self dual Ashtekar variables lead to an \textit{undeformed} algebra of the effective constraints, even when holonomy modifications are introduced \cite{SDSS}. Spherically symmetric LQG models, based on real variables, not only exhibits the same signature change deformation in the algebra of the modified constraints, but also the introduction of local physical degrees of freedom is forbidden in this model due to a partial no-go theorem presented in \cite{CovSpher}. Interestingly, it was shown in \cite{SDSS} that those two complexities go away when working with the self dual variables. It suggests therefore that the signature change in the deformed algebra approach could be a feature inherent to real variables. This distinctive property of self dual variables could be related to the well known fact that, contrary to the real Ashtekar-Barbero connection, the self dual one transforms as a true space-time connection \cite{SAMUEL} and perhaps is a better candidate to implement the classical symmetries of general relativity (GR) without deformations at the quantum level.

In this article, we show that the algebra of constraints for scalar cosmological perturbations, using self dual Ashtekar variables, remain \textit{undeformed}, on expected lines. This result thus hints at a general trend regarding the use of self dual variables in minisuperspace models of LQG and opens up possibilities of a new quantization scheme. Remarkably, the precise form of the polymerization function did not have to be fixed to make our case, thus allowing for quantization ambiguities. Although the precise form of the holonomy modification function is yet unknown for self-dual variables, (see \cite{JBACosmo, EdSDCosmo1, EdSDCosmo2} for some attempt to derive it), work on this is currently underway with some promising initial results. Nevertheless, several interesting novel features arise within this framework. While the algebra of quantum-corrected constraints remains the same as the classical one, we find as a necessary side-effect, that the spatial diffeomorphism constraint gets modified by some of these quantum correction functions and the gauge flow it generates is consequently more complicated. At first sight, this might look problematic since it is then not possible to implement our quantization using the characteristic spin network states. However, traditional spin networks can, anyway, not be used since the internal gauge group in this case is $SL(2,\mathbb{C})$ and not the usual $SU(2)$. Indeed, the construction of non compact spin networks remains challenging even today. (See \cite{NonCompactSN} for an attempt to define them in 2+1 gravity and \cite{PSN1, PSN2} for the introduction of the so called projected spin networks and \cite{NCSP2} for covariant twisted geometries.) On the other hand, modifications to spatial diffemorphisms suggest that the underlying spacetime might not be classical after all and gets deformed in an interesting manner. This has been discussed in some detail in our concluding remarks. Finally, the fact that no deformation occurs in our case, contrary to the deformed algebra approach using real variables, suggests that the predictions for the cosmic microwave background (CMB) power spectrum could be very different from this approach using self dual variables\footnote{Some `cyclic' interpretations in the deformed algebra approach for cosmological perturbations has been ruled out from phenomenological considerations \cite{DeformedAlgebraRuledOut}. We emphasize that it does not rule out the deformed algebra approach as a whole, let alone our reincarnation of it using self dual variables.}.

The paper has been organized as follows. We start with a lightning review of the real variables in LQC perturbations in the deformed algebra approach, emphasizing its salient features. Then we set up our calculations in the self dual case by deriving the (scalar) perturbation variables for cosmology around a flat FLRW background before going on to calculate the algebra of the (holonomy) modified constraints. This involves both the gravitational as well as matter contributions from a minimally coupled scalar field. Several restrictions on these modification functions follow from the requirement of the closure of the algebra, which are then shown to be consistent and yet giving rise to nontrivial solutions other than GR. Interpretations of our results and the possible next steps are finally discussed in the conclusion.

\section{Review of the real variables used in cosmological perturbations}
Cosmological perturbations have extensively been studied in the context of LQC using real-valued Ashtekar-Barbero variables (see \cite{Barrau:2013ula, Grain:2016jlq} for a review, and \cite{eucl3, LQCComparison2} for more conceptual discussions)\footnote{As mentioned in the introduction, we focus here only on the so-called deformed algebra of cosmological perturbations.}. At the background level first, i.e. strictly homogeneous and isotropic space-time, the connection variable and the densitized-triad variable simplify to $\bar{A}^i_a= c\delta^i_a$ and $\bar{E}_i^a=p\delta_i^a$ and their Poisson bracket is $\left\{c,p\right\}=\kappa\gamma/3V_0$, with $\gamma$ the real-valued Immirzi parameter, $\kappa=8\pi G$, and $V_0$ the fiducial spatial volume. The matter content is given by a scalar field, $\Psi$, with a Poisson bracket $\left\{\Psi,\pi\right\}=1/V_0$. In this symmetric case, the only non-vanishing constraint is the Hamiltonian constraint leading to $\bar{H}{\left[\bar{N}\right]}=\int_\Sigma d^3x \bar{N}\left[-3\sqrt{p}c^2/(\gamma^2\kappa)+\bar{\pi}/(2 p^{3/2})+ p^{3/2}V(\Psi)\right]$ where $V(\Psi)$ is the scalar field potential. The loop quantization of this symmetry-reduced system can be capture at an effective level by introducing the holonomy correction to the Hamiltonian constraint $c \to\sin(\delta c)/\delta$. With such quantum corrections, the cosmic history is drastically changed in the quantum regime since the big bang singularity is replaced by a regular quantum bounce\footnote{Indeed, in light of the findings of the inhomogeneous perturbations that we have a non-singular signature-change in the deep quantum regime, the simple picture of a bounce comes into question.} connecting a phase of classical contraction, to a phase of classical expansion (see e.g. \cite{lqc9, lqc10}).

To incorporate cosmological perturbations, one thus considers the perturbed FLRW space, meaning that the phase space variables are now ${A}^i_a=c \delta^i_a+\delta A^i_a$ and ${E}_i^a= p\delta_i^a+\delta E_i^a$ for the gravitational sector, and $\Psi=\bar{\Psi}+\delta\Psi$ and $\pi=\bar{\pi}+\delta\pi$ for the matter the sector. The barred quantities stands for the homogeneous and isotropic degrees of freedom while $\delta$-quantities are the perturbative quantities (not yet reduced to the gauge-invariant, physical degrees of freedom). Classically, the full dynamics is obtained by considering the Hamiltonian and diffeomorphism constraints expanded up to the quadratic order in perturbations. We note that in this expansion, the zeroth and second order of the diffeomorphism constraint are vanishing. The zeroth order Hamiltonian constraint gives the dynamics of the homogeneous and isotropic degrees of freedom (i.e. the dynamics of the strictly FLRW background). The first order Hamiltonian and diffeomorphism constraints generate the gauge transformations for the perturbations, allowing for defining gauge-invariant degrees of freedom. Finally, the second order Hamiltonian constraint generates the dynamics of the perturbations evolving in the FLRW background (see \cite{Langlois:1994ec} and \cite{Cailleteau:2011mi} for details of this approach in the ADM formalism and the Ashtekar formalism, respectively).

The dynamics of the cosmological perturbations now evolving in a background quantized using holonomies of the connection can be treated at an effective level by introducing quantum corrections replacing the background connection, $c$, by some function of it, $f(c)$, in the constraints expanded up to quadratic order. The perturbation part of the connection is not replaced by some function of $\delta A^i_a$. However, the background connection enters in the first and second order constraints thus affecting the dynamics of the perturbations. This function is $\sin(\delta c)/\delta$ for the zeroth order Hamiltonian constraints but they can a priori be different for the higher orders of the constraints. Since the matter sector of the constraints is independent of the connection $c$, only the gravitational constraints are a priori subject to the introduction of such quantum corrections. In LQG, one usually works with spin network states and the diffeomorphism constraints holds its classical form, meaning that a priori only the gravitational, Hamiltonian constraint receives quantum corrections. By doing that however, on ends up with a set of quantum corrected constraints which are not first class anymore: anomalous terms appear in the right-hand-side of the algebra of constraints given in the introduction (i.e. $\left\{ C_I,C_J \right\}=f^{K}_{~~IJ}(A^i_a,E_i^a)C_K$ which holds classically, is transformed into $\left\{ C^{qc}_I,C^{qc}_J\right\}=f^{K}_{~~IJ}(A^i_a,E_i^a)C^{qc}_K+A^{qc}_{IJ}$, with the superscript "$qc$" meaning quantum-corrected), thus violating covariance.

Quantum-corrected constraints which do form a set of first class constraints (that is cancelling the anomalous term) can nevertheless be retrieved, but this is made possible by allowing for the structure functions of the algebra to be also quantum-corrected, i.e. $\left\{ C_I,C_J\right\}=f^{K}_{~~IJ}(A^i_a,E_i^a)C_K$ to $\left\{ C^{qc}_I,C^{qc}_J\right\}={}^{qc}f^{K}_{~~IJ}(A^i_a,E_i^a)C^{qc}_K$ with ${}^{qc}f^K_{~~IJ}\neq f^{K}_{~~IJ}$, meaning that one ended with a {\it deformed} algebra as compared to general relativity. This is achieved by allowing for the matter part of the Hamiltonian constraint to be also affected by quantum corrections. More specifically in the case of a scalar field, it is sufficient to add quantum corrections to the spatial derivative term of the second order Hamiltonian of the matter, i.e. $\sqrt{p}\delta^{ab}(\partial_a\delta\Psi)(\partial_b\delta\Psi)\to \beta(c)\sqrt{p}\delta^{ab}(\partial_a\delta\Psi)(\partial_b\delta\Psi)$ \cite{tom1,tom2}. Remarkably enough, imposing these newly corrected constraints to form a first class algebra (in addition to impose that all the functions encoding quantum corrections tend to their GR form in the classical limit) unequivocally fixes all the introduced quantum corrections as functions of the background connection. In particular, $\beta(c)=\cos(2\delta c)$ which tends to one for small values of $\delta$. It is worth mentioning that not all the functions encoding the quantum corrections are arbitrarily chosen prior to imposing the closure of the algebra: the function for the zeroth-order Hamiltonian constraint (i.e. the background Hamiltonian) is set equal to the form obtained from the loop quantization, i.e. $\sin(\delta c)/\delta$. Other remarkable features of this approach are that the consistency relations that the quantum corrections should satisfy to ensure an anomaly-free algebra also fixes the quantization scheme of the background to be the so-called $\bar{\mu}$-scheme \cite{tom2}, and also imposes the diffeomorphism constraint to hold its GR form \cite{tom2}.

The deformation of the constraints algebra is completely encoded in the $\left\{H,H\right\}$ Poisson bracket, which now reads \cite{eucl2}
\begin{equation}	\left\{H^{qc}\left[N_1\right]\,,\,H^{qc}\left[N_2\right]\right\} = D^{qc}\left[\beta(c)q^{ab}\,\left(N_1\nabla_b N_2 - N_2\nabla_b N_1\right)\right].
\end{equation}
We stress that the modification of the algebra is solely given by the quantum correction introduced in the matter sector in this setting. The appearance of the $\beta$ function in this bracket has dramatic consequences from both the theoretical viewpoint, and the phenomenological viewpoint. From the theoretical viewpoint first, the algebra of constraints in GR is a representation of the algebra of the deformation of the hypersurfaces embedded in a (pseudo)Riemaniann space \cite{HOJMAN197688}. If the deformed algebra as obtained by implementing quantum corrections is to be representative of the hypersurface deformation of an underlying 4-dimensional structure, this would mean that this structure is not anymore given by (pseudo)Riemanian manifolds. In addition, the right-hand-side of the above can be written as $\beta(c)D^{qc}\left[q^{ab}\,\left(N_1\nabla_b N_2 - N_2\nabla_b N_1\right)\right]$ since $c$ is an homogeneous degree of freedom. In GR, the $\left\{H,H\right\}$ Poisson bracket is equal to $sD^{qc}\left[q^{ab}\,\left(N_1\nabla_b N_2 - N_2\nabla_b N_1\right)\right]$, with $s=1$ for the Lorentzian signature, and $s=-1$ for the Euclidean signature \cite{lqg6}. For $c\in[\pi/4\delta,3\pi/4\delta]$ which corresponds to the quantum regime of the cosmic history (i.e. close to the quantum bounce), the function $\beta$ becomes negative-valued, while it is positive for other values of $c$ corresponding to either the classical contraction or the classical expansion. If the sign of $\beta$ is interpreted as giving the signature of the underlying 4-dimensional structure, this would mean that the quantum-corrected space-time can experienced signature change \cite{Bojowald:2011aa, Bojowald:2012ux} (which points again into the fact that this underlying 4-dimensional structure cannot be (pseudo)Riemaniann manifolds). This astonishing feature raises some difficulties if one is willing to reconstruct a 4-dimensional structure on which the above-derived deformed algebra of hypersurface deformation could hold.

From the phenomenological viewpoint then, the presence of $\beta$ in the matter sector which can become negative-valued means that cosmological perturbations will experienced kinds of tachyonic instabilities around the bounce \cite{JulienSignChange, MielSignChange, MielAsymSilence}. This can be intuitively understood as follows. Working in the spatially flat gauge for cosmological perturbations, the gauge-invariant scalar degree of freedom then co\"\i{n}cides with the scalar field perturbations. Because the gradient part of the scalar field Hamiltonian at second order is now given by $\beta(c)\sqrt{p}\delta(\partial_a\delta\Psi)(\partial_b\delta\Psi)$, one easily figures out that the spatial-derivative part of the equation of  motion for scalar perturbations will be $\propto\beta(c)\partial_a\partial^a\delta\Psi$, where the instability develops for $\beta<0$ (note that the same instability also applies to the tensor modes, see \cite{eucl2}). Depending on the specific choice to set the initial conditions for the perturbations, this in particular leads to dramatic departures from the nearly scale-invariant power spectrum of cosmological perturbations \cite{Schander:2015eja, Bolliet:2015bka, Bolliet:2015raa} as now very well constrained from CMB datas \cite{Planck1, Planck2} (see also \cite{Bojowald:2015gra, Mielczarek:2014kea} for alternatives).

We stress that the above discussion assumes that only holonomy corrections are taken into account. A similar approach has been developed for inverse-triad corrections in Ref. \cite{Bojowald:2008jv}, and for the joint case of inverse-triad and holonomy corrections in Ref. \cite{Cailleteau:2013kqa}. In the latter, there exists more than one single solution to the consistency relations the quantum corrections should satisfy to get an anomaly free algebra, and in some of these solutions, the signature-change disappears.

\section{The preliminaries}
In term of the self dual Ashtekar variables, the phase space of GR is given by the following Poisson bracket
\be
\{ A^i_a, E^b_j\} = i \delta^b_a \delta^i_j\,,
\ee
constrained to satisfy
\begin{align}
& D_{\text{grav}}[N^a] = \frac{1}{i\kappa} \int_{\Sigma} dx^3 \; N^a \; [ \; F^i_{ab} E^b_i - A^i_a \cG_i \; ], \\
& H_{\text{grav}}[N] = \frac{1}{2\kappa} \int_{\Sigma} dx^3 N \; \frac{E^a_i E^b_j}{\sqrt{|E|}} \; \epsilon^{ij}{}_k F^k_{ab}.  %\approx 0\,, \qquad D_{\text{grav}}[N^a] = \frac{1}{i\kappa} \int_{\Sigma} dx^3 \; N^a \; [ \; F^i_{ab} E^b_i - A^i_a \cG_i \; ] \approx 0\,.
\end{align}
In the above, $F^i_{ab}=\partial_a A^i_b-\partial_b A^i_a+\epsilon^i_{jk}A^j_aA^k_b$ is the curvature of the Ashtekar connection, and $E=\det{E^a_i}$.

We symmetry reduce the system to that of a homogeneous, isotropic cosmology as given by a FLRW geometry, where we have only a background connection and a densitized triad, denoted by $c$ and $p$ respectively \cite{lqc10}. Defining a fiducial volume to get rid of the difficulties of having integrals of infinite (spatial) volumes, $V_0$, the Poisson bracket on the reduced phase space takes the form
\bea
\{c, p\} = \frac{i\kappa}{3V_0}\,,
\eea
while the constraints are derived in the next section.

Adding a minimally coupled scalar field $\Psi$ to the system, we obtain the following extension of the phase space
\be
\{ \Psi, \pi\} = \frac{1}{V_0}\,,
\ee
where the scalar Hamiltonian and diffeomorphism constraints are given by
\bea
H_{\text{scal}}[N] &=& \int_{\Sigma} dx^3 \; N\; [ \; \frac{\pi^2}{2 \sqrt{|E|}} + \frac{1}{2\sqrt{|E|}} E^a_i E^{bi} \partial_a \Psi \partial_b \Psi + \sqrt{|E|} V(\Psi) \; ]\,,\\
D_{\text{scal}}[N^a] &=& \int_{\Sigma} dx^3 \; N^a \; \pi \partial_a \Psi\,.
\eea
with a potential $V=V(\Psi)$ for the scalar field.

\subsection{Reducing to a perturbed FLRW  geometry}
The starting point is to consider a perturbed FLRW metric given by
\be
g_{\mu\nu} = a^2 \mat{cccc}{-1 & 0 \\ 0 & \delta_{ab}} + a^2 \mat{cccc}{- 2 \phi & \partial_a B \\\partial_b B & - 2 \psi \delta_{ab} + \partial_a \partial_b E} + a^2 \mat{cccc}{0 & S_a \\ S_b & \nabla_{[a}F_{b]}} + a^2 \mat{cccc}{0 & 0 \\ 0 & h_{ab}}\,.
\ee
The first matrix correspond to the background geometry, the usual FLRW universe, while the three others represent respectively the scalar, vectorial and tensorial perturbations to the background geometry. The Scalar-Vector-Tensor decomposition of perturbations must satisfy the following conditions
\be
\partial^{a} h_{ab} = 0\,, \qquad \delta^{ab} h_{ab} = 0\,, \qquad \nabla_aF^a = 0\,, \qquad \nabla_aS^a = 0\,.
\ee
In this paper, we shall only be concerned with scalar perturbations since it is the most complicated one to handle, and leave the other ones for future work.

Reducing the full phase space of GR (coupled to a scalar field) to the one of the perturbed FLRW geometry (coupled to a scalar field), we end up with the following variable (where the bar denote the background nature of the variable):
\be
A^i_a = \bar{A}^i_a  + \delta A^i_a\,, \qquad E^a_i = \bar{E}^a_i + \delta E^a_i\,, \qquad \Psi = \bar{\Psi} + \delta \Psi\,, \qquad \pi = \bar{\pi} + \delta \pi\,,
\ee
where
\be
\bar{A}^i_a = c \delta^i_a\,,  \qquad \bar{E}^a_i = p \delta^a_i\,, \qquad \text{and} \qquad \{ c , p\} = \frac{i\kappa}{3V_0}\,,
\ee
while the Poisson bracket for the perturbed quantities are given by
\bea
\{ \delta A^i_a(x), \delta E^b_j (y)\} = i\kappa \delta^a_b \delta^j_i \delta^3(x,y)\,.
\eea

The perturbative expansion to quadratic order of the diffeomorphism and Hamiltonian constraints involves such terms as $F^i_{ab}$, $E^a_iE^b_j$, and $E^{\pm 1/2}$. They are given by
\bea
F^i_{ab}  = c^2\epsilon^i{}_{jk} \delta^j_a \delta^k_b + (\partial_a \delta A^i_b - \partial_b \delta A^i_a + c\epsilon^i{}_{jk} ( \delta A^j_a \delta^k_b  + \delta^j_a \delta A^k_b) ) + \epsilon^i{}_{jk}  \delta A^j_a \delta A^k_b\,,
\eea
\bea
E^a_i E^b_j  = p^2 \delta^a_i \delta^b_j + p ( \delta^a_i \delta E^b_j + \delta^b_j \delta E^a_i) + \delta E^a_i \delta E^b_j\,,
\eea
\bea
\label{E}
 |E|^{\pm 1/2} = p^{\pm 3/2} \; [ \; 1   \pm \frac{1}{2p} \delta E^a_i\delta^i_a \mp \frac{1}{4 p^2}  \delta^i_a \delta E^a_j \delta^j_b \delta E^b_i  + \frac{1}{8 p^2}   (\delta^i_a \delta E^a_i)^2 + \mathcal{O}(E^3)  \; ]\,.
\eea
The last expansion, Eq. (\ref{E}), is derived from $$\det(M)=\frac{1}{6}\left[(\mathrm{Tr}(M))^3-3\mathrm{Tr}(M)\mathrm{Tr}(M^2)+2\mathrm{Tr}(M^3)\right],$$ with $M$ a $(3\times3)$-square matrix. (This is obtained as a consequence of the Cayley-Hamilton theorem applied to $(3\times3)$-square matrix.) Having listed the expansion of the most non trivial terms, we can now compute explicitly the expression of the perturbed constraints.

\subsection{The perturbed constraints}

\subsubsection{Gravitational perturbed scalar constraint}

The (unsmeared) self dual gravitational Hamiltonian constraint is given by

\begin{align*}
\mathcal{H}_{\text{grav}}& = \; \frac{E^a_i E^b_j}{\sqrt{|E|}} \; \epsilon^{ij}{}_k F^k_{ab}\,.
%& = \; p^{- 3/2} \; [ \; 1  - \frac{1}{2p} \delta^m_c \delta E^c_m + \frac{1}{4 p^2}  \delta^m_c \delta E^c_n \delta^n_d\delta E^d_m  +  \frac{1}{8 p^2} (\delta^m_c\delta E^c_m)^2 \; ] \\
%& \hspace{1cm}\; \times [ \; p^2 \delta^a_i \delta^b_j + p ( \delta^a_i \delta E^b_j + \delta^b_j \delta E^a_i) + \delta E^a_i \delta E^b_j \; ] \\
%& \hspace{1cm}\; \times \; \epsilon^{ij}{}_k [ \; c^2\epsilon^k{}_{mn} \delta^m_a \delta^n_b  + (\partial_a \delta A^k_b - \partial_b \delta A^k_a + c\epsilon^k{}_{mn} ( \delta A^m_a \delta^n_b  + \delta^m_a \delta A^n_b) )  + \epsilon^k{}_{mn}  \delta A^m_a \delta A^n_b \; ]
\end{align*}
Proceeding to its expansion up to quadratic order, we have
\begin{align*}
\mathcal{H}_{\text{grav}} %&  = \; p^{- 3/2} \; [ \; p^2 \delta^a_i \delta^b_j \; ( \; 1  - \frac{1}{2p} \delta E + \frac{1}{4 p^2} \delta E^c_q \delta E^q_c  +  \frac{1}{8 p^2} \delta E^2 \; ) + p ( \delta^a_i \delta E^b_j + \delta^b_j \delta E^a_i) \; (\; 1  - \frac{1}{2p} \delta E  \;)  + \delta E^a_i \delta E^b_j \; ] \; \\
%& \times \; \epsilon^{ij}{}_k [ \; \epsilon^k{}_{mn} \delta^m_a \delta^n_b \bar{k}^2  + (\partial_a \delta A^k_b - \partial_b \delta A^k_a + \epsilon^k{}_{mn} ( \delta A^m_a \delta^n_b  + \delta^m_a \delta A^n_b) ) \bar{k} + \epsilon^k{}_{mn}  \delta A^m_a \delta A^n_b \; ] \\ \\
& = \; [ \; p^{1/2} \delta^a_i \delta^b_j  - \frac{1}{2} \; p^{-1/2} \delta^a_i \delta^b_j \delta E + p^{-1/2}  ( \delta^a_i \delta E^b_j + \delta^b_j \delta E^a_i) + \frac{1}{4} \; p^{-3/2} \delta^a_i \delta^b_j \;\delta E^q_c \delta E^c_q  + \frac{1}{8} \; p^{-3/2} \delta^a_i \delta^b_j  \; \delta E^2 \\
& - \frac{1}{2} \; p^{-3/2} \;  \delta E \; ( \delta^a_i \delta E^b_j + \delta^b_j \delta E^a_i) + p^{-3/2} \delta E^a_i \delta E^b_j \; ] \\
& \times \; \epsilon^{ij}{}_k [ \; \epsilon^k{}_{mn} \delta^m_a \delta^n_b {c}^2  + (\partial_a \delta A^k_b - \partial_b \delta A^k_a + \epsilon^k{}_{mn} ( \delta A^m_a \delta^n_b  + \delta^m_a \delta A^n_b) ) {c} + \epsilon^k{}_{mn}  \delta A^m_a \delta A^n_b \; ]\,.
\end{align*}
Regrouping the different order involved in the Hamiltonian constraint, we end up with
\begin{align}
& \mathcal{H}^{(0)}_{\text{grav}}  = 6 \sqrt{p} c^2 \\
& \mathcal{H}^{(1)}_{\text{grav}} = 2 \sqrt{p} \; \epsilon^{ab}{}_k  \; \partial_a \delta A^k_b + 4 c \sqrt{p} \;  \delta A^i_a \delta^a_i  +  \frac{1}{ \sqrt{p}} \; c^2 \delta E^a_i \delta^i_a \\
& \mathcal{H}^{(2)}_{\text{grav}}  = \frac{1}{2} p^{-3/2} c^2 \; \delta^i_a\delta E^a_j\delta^j_b \delta E^b_i - \frac{1}{4} p^{-3/2} c^2 (\delta^i_a\delta E^a_i)^2  + \sqrt{p} \; (\delta^a_i\delta A^i_a)^2 - \sqrt{p} \; \delta A^i_a \delta^a_j\delta A^j_b \delta^b_i + \frac{2}{\sqrt{p}} \; c \; \delta E^a_i \delta A^a_i \nn \\
& \qquad \qquad - \frac{1}{\sqrt{p}}  \; (\delta^j_c\delta E^c_j) \;  \epsilon^{ab}{}_k \; \partial_a \delta A^k_b + \frac{2}{\sqrt{p}} \; \delta^a_i \delta E^b_j \; \epsilon^{ij}{}_k \; \partial_{[a} \delta A^k_{b]}\,.
\end{align}
The full gravitational constraint, up to second order in perturbations, can thus be written as
\bea
H_{\text{grav}}[N] &=& \frac{1}{2 \kappa} \int \text{d}^3 x \left(\bar{N} \left[\mathcal{H}^{(0)}_{\text{grav}} + \mathcal{H}^{(2)}_{\text{grav}}\right] + \delta N \mathcal{H}^{(1)}_{\text{grav}}\right)\,,
\eea
where the lapse function has been split as $N(x,t) = \bar{N}(t) +\delta N(x,t)$.

\subsubsection{Gravitational perturbed diffeomorphism constraint}

Classically solving for the Gauss constraint, one can expand the diffeomorphism constraint as
\begin{align*}
 \mathcal{D}_{\text{grav}}& = \; F^i_{ab} E^b_i  \\
 %& = \; [ \; p \delta^b_i + \delta E^b_i \; ]  \times [ \; \epsilon^i{}_{jk} \bar{A}^j_a \bar{A}^k_b + (\partial_a \delta A^i_b - \partial_b \delta A^i_a + \epsilon^i{}_{jk} ( \delta A^j_a \bar{A}^k_b  + \bar{A}^j_a \delta A^k_b) ) + \epsilon^i{}_{jk}  \delta A^j_a \delta A^k_b \; ] \\
 & =  \; p \delta^b_i  \times [ \; \epsilon^i{}_{jk} \delta^j_a \delta^k_b {c}^2+ (\partial_a \delta A^i_b - \partial_b \delta A^i_a + \epsilon^i{}_{jk} ( \delta A^j_a \delta^k_b  + \delta^j_a \delta A^k_b) ) {c} + \epsilon^i{}_{jk}  \delta A^j_a \delta A^k_b \; ] \\
 & +  \;  \delta E^b_i \; \times [ \; \epsilon^i{}_{jk} \delta^j_a \delta^k_b {c}^2 + (\partial_a \delta A^i_b - \partial_b \delta A^i_a + \epsilon^i{}_{jk} ( \delta A^j_a \delta^k_b  + \delta^j_a \delta A^k_b) ) {c}\; ]\,.
\end{align*}
Regrouping the different order involved in the diffeomorphism constraint, we have
\begin{align}
 %& \mathcal{D}^{(0)}_{\text{grav}} = p c^2 \delta_b . (\delta_a \times \delta_b) = 0 \\
 & \mathcal{D}^{(0)}_{\text{grav}} = p c^2 \delta^j_a\delta^k_i\epsilon^i{}_{jk} = 0 \\
 & \mathcal{D}^{(1)}_{\text{grav}} = p \; \partial_a \delta A - p \delta^b_i \partial_b \delta A^i_a + p c \;  \epsilon^b{}_{ak} \;  \delta A^k_b + c^2 \epsilon^i{}_{ab} \delta E^b_i \\
  & \mathcal{D}^{(2)}_{\text{grav}} = p \;  \delta^b_i \epsilon^i{}_{jk} \delta A^j_a \delta A^k_b + \delta E^b_i \partial_{[a} \delta A^i_{b]}\,.
\end{align}
Keeping in mind that the background shift vector is trivially zero, we have $N^a(x,t) = \delta N^a(x,t)$. We can write the full diffeomorphism constraint, up to the second order in perturbation, as
\bea
D_{\text{grav}}[N^a]=\frac{1}{i\kappa} \int \text{d}^3 x \;  \delta N^a \;  \mathcal{D}^{(1)}_{\text{grav}}\,.
\eea

\subsubsection{Perturbed Hamiltonian constraint for the scalar field}

Next we compute the contribution of the scalar field  to the (unsmeared) Hamiltonian constraint which becomes
\begin{align*}
\mathcal{H}_{\text{scal}} & = \; \frac{\pi^2}{2 \sqrt{|E|}} + \frac{1}{2\sqrt{|E|}} E^a_i E^{bi} \partial_a \Psi \partial_b \Psi + \sqrt{|E|} V(\Psi) \;  \\
%& = \; p^{-3/2}[ \; \frac{1}{2}  - \frac{1}{4p} \delta^i_a \delta E^a_i + \frac{1}{8 p^2}  \delta^i_a \delta E^a_j \delta E^j_b \delta^b_i +  \frac{1}{16 p^2} \delta^i_a  \delta E^a_i  \delta^j_b \delta E^b_j ] [ (\bar{\pi} + \delta \pi)^2 + E^a_i E^{bi} \partial_a (\bar{\Psi} + \delta \Psi) \partial_b (\bar{\Psi} + \delta \Psi)  ] \\
%& +  \; p^{ 3/2} \; [ \; 1  + \frac{1}{2p} \delta^i_a \delta E^a_i - \frac{1}{8 p^2}  \delta^i_a \delta E^a_j \delta E^j_b \delta^b_i -  \frac{1}{4 p^2}  \delta E^a_i \delta E^i_a \; ]  [  V(\bar{\Psi}) + \frac{\delta V}{\delta \bar{\Psi}} \delta \Psi \; ] \\
& = \; p^{-3/2}[ \; \frac{1}{2}  - \frac{1}{4p} \delta^i_a \delta E^a_i + \frac{1}{8 p^2}  \delta^i_a \delta E^a_j \delta E^j_b \delta^b_i +  \frac{1}{16 p^2} \delta^i_a  \delta E^a_i  \delta^j_b \delta E^b_j ] [ (\bar{\pi} + \delta \pi)^2 + E^a_i E^{bi} \partial_a  \delta \Psi \partial_b  \delta \Psi  ] \\
& +  \; p^{ 3/2} \; [ \; 1  + \frac{1}{2p} \delta^i_a \delta E^a_i - \frac{1}{8 p^2}  \delta^i_a \delta E^a_j \delta E^j_b \delta^b_i -  \frac{1}{4 p^2}  \delta E^a_i \delta E^i_a \; ]  [  V(\bar{\Psi}) + \frac{\delta V}{\delta \bar{\Psi}} \delta \Psi + \frac{\delta^2 V}{\delta \bar{\Psi}^2} (\delta \Psi)^2  \; ]\,.
\end{align*}
Isolating the zero, first and second order, we obtain
\begin{align}
\mathcal{H}^{(0)}_{\text{scal}} &  = \frac{\bar{\pi}^2}{2p^{3/2}}  + p^{ 3/2}  V(\bar{\Psi}) \\
\mathcal{H}^{(1)}_{\text{scal}} &  = \frac{\pi \delta \pi}{p^{3/2}} - \frac{\pi^2}{2 p^{3/2}} \frac{\left(\delta E^c_j\delta^j_c\right)}{2 p} + p^{3/2} \; [ \; V_{, \Psi} \delta \psi + V \; \frac{\left(\delta E^c_j\delta^j_c\right)}{2p} \; ] \\
\mathcal{H}^{(2)}_{\text{scal}} &  = \frac{\delta \pi^2}{2 p^{3/2}} - \frac{\pi \delta \pi}{p^{3/2}} \frac{\delta E}{2p} + \frac{\pi^2}{2 p^{3/2}} \; [ \;  \frac{\left(\delta E^c_j\delta^j_c\right)^2}{8 p^2} + \frac{1}{4 p^2} \delta^k_c \delta^j_d \delta E^c_j \delta E^d_k \; ] + \frac{\sqrt{p}}{2}  \delta^{ab} \partial_a \delta \Psi \partial_b \delta \Psi + \frac{p^{3/2}}{2} V_{, \Psi \Psi} \delta \Psi^2  \nn \\
& + p^{3/2} V_{, \Psi} \delta \Psi \frac{\delta E}{2p}  + p^{3/2} V \; [ \; \frac{\left(\delta E^c_j\delta^j_c\right)^2}{8 p^2} - \frac{1}{4 p^2} \delta^k_c \delta^j_d \delta E^c_j \delta E^d_k \; ]\,.
\end{align}
The full gravitational constraint, up to second order in perturbations, can thus be written as
\bea
H_{\text{scal}}[N] &=& \frac{1}{2 \kappa} \int \text{d}^3 x \left(\bar{N} \left[\mathcal{H}^{(0)}_{\text{scal}} + \mathcal{H}^{(2)}_{\text{scal}}\right] + \delta N \mathcal{H}^{(1)}_{\text{scal}}\right)\,.
\eea

\subsubsection{ Perturbed diffeomorphism constraint for the scalar field}

We complete the list by the perturbed (unsmeared) diffeomorphism constraint for the scalar field
\begin{align*}
 \mathcal{D}_{\text{scal}} & = \pi \partial_a \Psi \\
 & = ( \bar{\pi} + \delta \pi ) \partial_a ( \bar{\Psi} + \delta \Psi)  \\
 & = \; \bar{\pi} \partial_a  \bar{\Psi} + \bar{\pi} \partial_a \delta \Psi  +  \delta \pi  \partial_a  \bar{\Psi} +  \delta \pi  \partial_a \delta \Psi\,,
\end{align*}
where we have
\begin{align}
  \mathcal{D}^{(0)}_{\text{scal}}  = \; \bar{\pi} \partial_a  \bar{\Psi} = 0  \qquad  \mathcal{D}^{(1)}_{\text{scal}}  =  \; \bar{\pi} \partial_a \delta \Psi   \;   \qquad \mathcal{D}^{(2)}_{\text{scal}} =  \delta \pi  \partial_a \delta \Psi\,.
\end{align}
The full scalar diffeomorphism constraint, up to the second order in perturbation, as
\bea
D_{\text{scal}}[N^a]=\frac{1}{i\kappa} \int \text{d}^3 x \; \delta N^a \;  \mathcal{D}^{(1)}_{\text{scal}}\,.
\eea

\section{Implementing holonomy corrections in the effective constraints}
Let us now implement the quantum corrections as one usually introduces in effective LQC \cite{LQGEffectiveTheory1, LQGEffectiveTheory2}. The quantum corrections will only affect the background gravitational canonical variables, and we restrict to the so called holonomy corrections, i.e. the corrections affecting $c$, disregarding the inverse triad corrections. The quantum corrections implemented here follows from the fact that one regularizes the Hamiltonian as well as the diffeomorphism constraint so that those constraints are expressed in term of the holonomy of the connections $h_{\mu}(\bar{A})$ and no more directly in term of the connection $\bar{A}^i_a = c \delta^i_a$.

At the effective level, this implies that each function $c(t)$ is replaced by a function $f(c(t))$. In real LQC, this function is given by the $SU(2)$ group structure and chosen to be $\sin{(\delta c)}/ \delta$. Here we keep this function general, and we simply implement the following general quantum correction
\be
c \; \rightarrow \;  f(c)\,,
\ee
introducing a different function for each $c$ appearing in the constraints. The philosophy we adopt is that we introduce the least amount of arbitrary functions necessary to get a non-trivial implementation of holonomy corrections in the end. As we shall see, requiring that there are no anomalies in the full hypersurface deformation algebra would lead to several restrictions on these arbitrary functions. A priori, there is no reason to believe that these consistency conditions would not be severe enough to restrict these correction functions to their trivial classical values, as has happened (partially) in some cases in the past \cite{BojowaldShanky}. Thus, although we begin with a large number of arbitrary functions which would allow for various quantization ambiguities, there are a much more general ansatz which we could have chosen by simply requiring that some of these correction functions go to unity in the classical limit\footnote{The point of our argument can be illustrated by the appearance of the function $h_4(c)$, which is required to go to $1$ in the classical limit, in two places.}. We simply choose to work with the least amount of such modifications as necessary to consistently implement holonomy corrections for cosmological perturbations.

The gravitational scalar constraint becomes, at each order
\bea\label{modHam}
\mathcal{H}^{(0)}_{\text{grav}} &  =  &6 \sqrt{p} \; g (c) \\
\mathcal{H}^{(1)}_{\text{grav}}
& = &2 \sqrt{p} \; \epsilon^{ab}{}_k  \; \partial_a \delta A^k_b + 4 \sqrt{p} f_1(c) \;  \delta A^i_a\delta^a_i  +  \frac{1}{ \sqrt{p}} \; f_2 (c) \delta E^a_i\delta^i_a \\
\mathcal{H}^{(2)}_{\text{grav}}
& = &\frac{1}{2} p^{-3/2} h_1 (c) \; \delta E^a_i\delta^i_b \delta E^b_j \delta^j_a - \frac{1}{4} p^{-3/2} h_2(c) (\delta E^a_i \delta^i_a)^2  + \sqrt{p}\;h_4(c) \; (\delta A^i_a\delta^a_i)^2 - \sqrt{p}\; h_4(c) \; \delta A^i_a\delta^a_j \delta A^j_b \delta^b_i \nn  \\
&  &\; + \frac{2}{\sqrt{p}} \; h_3(c) \; \delta E^a_i \delta A^i_a - \frac{1}{\sqrt{p}}  \; (\delta E^a_i\delta^i_a) \;  \epsilon^{ab}{}_k \; \partial_a \delta A^k_b + \frac{2}{\sqrt{p}} \; \delta^a_i \delta E^b_j \; \epsilon^{ij}{}_k \; \partial_{[a} \delta A^k_{b]}\,,
\eea
while the modified diffeomorphism constraint takes the form
\bea\label{modDiffeo}
  D^{(1)}_{\text{grav}} & = p \; \partial_a \delta A - p \delta^b_i \partial_b \delta A^i_a + p \tilde{f}_1(c) \;  \epsilon^b{}_{ak} \;  \delta A^k_b + \tilde{f}_2(c) \epsilon^i{}_{ab} \delta E^b_i\,.
\eea
Including holonomy correction functions in the diffeomorphism constraint deserves some justification in this case. In LQG, one typically works with spin network states and thereby impose the diffeomorphism constraint classically and impose the quantum Hamiltonian constraint on diff-invariant states. However, the spin network states are, after all, a choice of basis and, therefore, it is entirely possible to work in some representation whereby one needs an infinitesimal quantum operator corresponding to the diffeomorphism constraint. Indeed we are asking questions regarding the algebra of quantum constraints in this work, which should be, to a large extent, a representation independent question. What this calculation perhaps suggests is that the flow of the diffeomorphsim constraint is crucially different from the classical flow, in the higher curvature regime, as is the case for the Hamiltonian one\footnote{For an attempt to define an infinitesimal diffeomorphism constraint in LQG, see \cite{Varadarajan} and references therein.}. Thus we keep this possibility of modification in our calculations.

The constraint related to the scalar field are not modified, since the connection $A^i_a$ is not involved in its expression. This is a major point of departure in our formalism from the real-valued variables, where one is forced to introduce counter-terms in the matter part as well.

\section{Anomaly free algebra of the (self dual) gravitational effective phase space}

We begin by showing that the hypersurface deformation algebra remains anomaly free and even undeformed for the gravitational sector, although we modify the constraints through holonomy corrections as mentioned above.

\subsection{The $\{ H_{\text{grav}}, H_{\text{grav}}\}$ bracket }

Let us first compute the Poisson bracket involving the scalar constraint with itself. Since we are dealing with the gravitational sector first, we drop the subscript 'grav' in the following notation. Because of homogeneity and isotropy, the two lapse functions $N_1$ and $N_2$ involved in the Poisson bracket  $\{ H[N_1], H[N_2]\}$ can be written as $N_i (t,x)= \bar{N}_i(t) + \delta N_i (t,x)$ (the subscript $i$ here means that the lapse is evaluated at different spatial points, not to be understood as a internal Lorentz index).
Therefore, the Poisson bracket reduces to
\begin{align*}
\{ H[N_1], H[N_2]\} & = \{ H^{(0)}[\bar{N}_1] + H^{(1)}[\delta N_1] + H^{(2)}[\bar{N}_1] \; , \;  H^{(0)}[\bar{N}_2] + H^{(1)}[\delta N_2] + H^{(2)}[\bar{N}_2] \} \\
& = \{  H^{(0)}[\bar{N}_1] \; , \;  H^{(0)}[\bar{N}_2]  \} +\{  H^{(0)}[\bar{N}_1] \; , \;  H^{(1)}[\delta N_2] \} +  \{  H^{(0)}[ \bar{N}_1] \; , \;  H^{(2)}[\bar{N}_2] \}\\
& +  \{  H^{(1)}[\delta N_1] \; , \;  H^{(0)}[\bar{N}_2]  \} +\{  H^{(1)}[\delta N_1] \; , \;  H^{(1)}[\delta N_2] \} +  \{  H^{(1)}[ \delta N_1] \; , \;  H^{(2)}[\bar{N}_2] \}\\
& +  \{  H^{(2)}[\bar{N}_1] \; , \;  H^{(0)}[\bar{N}_2]  \} +\{  H^{(2)}[\bar{N}_1] \; , \;  H^{(1)}[\delta N_2] \} +  \{  H^{(2)}[ \bar{N}_1] \; , \;  H^{(2)}[\bar{N}_2] \}\\
\end{align*}
Now, because of homogeneity and isotropy, and considering for example only the scalar perturbations, the background lapse reads $\bar{N}_i (t) = \bar{N} = a(t)$, and we can therefore remove its label, while the perturbed lapse is given by $\delta N_i (t,x) = 2 a(t) \phi (t,x)$ and one has to keep its label since it depends on space (note that any term in the brackets involving $\partial_a \bar{N} $ vanishes).

Keeping this in mind, it is obvious that the following terms commute with themselves
\begin{align*}
\{  H^{(0)}[\bar{N}] \; , \;  H^{(0)}[\bar{N}]  \} & = \{ H^{(2)}[\bar{N}], H^{(2)} [\bar{N}]\} = 0\,.
\end{align*}

The following two terms cancel each other
\bea
\{  H^{(0)}[\bar{N}] \; , \;  H^{(2)}[\bar{N}] \} + \{  H^{(2)}[\bar{N}] \; , \;  H^{(0)}[\bar{N}]  \} = 0
\eea
since the resulting terms are proportional to $\bar{N}^2$ (which gets cancelled by a contribution with an opposite sign from the second bracket by symmetry of the Poisson bracket structure) and because there are no terms with derivatives on the background variables and therefore no integration by part involved in those brackets.

%Indeed, we have obviously that $\{  H^{(0)}[\bar{N}_1] \; , \;  H^{(0)}[\bar{N}_2]  \} = \{  H^{(2)}[\bar{N}_1] \; , \;  H^{(2)}[\bar{N}_2]  \} = 0$, as well as $ \{  H^{(0)}[\bar{N}_1] \; , \;  H^{(2)}[\bar{N}_2]  \} + \{  H^{(2)}[\bar{N}_1] \; , \;  H^{(0)}[\bar{N}_2]  \}  = 0$. Finally, we can show moreover that
%\begin{align*}
% \{  H^{(0)}[\bar{N}_1] \; , \;  H^{(1)}[\delta N_2]  \} + \{  H^{(1)}[\delta N_1] \; , \;  H^{(0)}[\bar{N}_2]  \} & = 0
%\end{align*}

We compute the non trivial bracket
\bea
& &\{  H^{(1)}[\delta N_1] \; , \;  H^{(1)}[\delta N_2]  \}  \nn\\
& &=  \frac{1}{4\kappa^2}\int dx^3 dy^3  \{ \delta N_1 (x) ( 2 \sqrt{p} \epsilon^{ab}{}_k \partial_a \delta A^k_b(x) + 4 \sqrt{p} f_1(c) \delta A^i_a(x) \delta^a_i + \frac{f_2(c)}{\sqrt{p}} \delta E^a_i(x) \delta^i_a)  \; ,\nn\\
& &\hspace{2cm}  \delta N_2 (y) ( 2 \sqrt{p} \epsilon^{cd}{}_i \partial_c \delta A^i_d  + 4 \sqrt{p} f_1(c) \delta A^m_c(y)\delta^c_m + \frac{f_2(c)}{\sqrt{p}} \delta E^c_m(y) \delta^m_c)  \; \} \nn\\
& &= \frac{i}{4\kappa}\int dx^3 \; \left[ \delta N_1(x) \delta N_2(x) - \delta N_2(x) \delta N_1(x) \right] \left[\cdots\right] \nn\\
& &\hspace{2cm} + \frac{1}{\kappa} \int dx^3 dy^3 (\delta N_1 (x) \; \epsilon^{ab}{}_k \partial_a \delta A^k_b (x) \; , \; \delta N_2(y) \;f_2(c) \delta^c_m \delta E^m_c (y) \} + \text{sym} \nn\\
& &=  \frac{i}{\kappa} \int dx^3 dy^3 \; \left[N_1 \partial_a \delta N_2 - N_2 \partial_a \delta N_1\right] \; f_2(c) \;\epsilon^{ab}{}_k \delta^k_b  = 0
\eea
The first term of the second equality above comes from the Poisson bracket between the background variables and thus there are no derivative terms to integrate by parts and thus they are cancelled by a same term with opposite sign due to the anti-symmetry of Poisson brackets. On the other hand, the term in the last equality goes to zero due to its tensorial structure, i.e. $\epsilon^{ab}{}_k \delta^k_b=0$.
%For the very last term, we can already conclude without doing the computation that it vanishes. Indeed, there will be two kind of terms, 1) terms of third order that we can therefore simply remove even if there are not vanishing, 2) terms of second order. in this class, there will be terms which

Therefore, we are left with the following terms
\begin{align}
\{ H[N_1], H[N_2]\}
& = \{  H^{(0)}[\bar{N}] \; , \;  H^{(1)}[\delta N_2] \}  +  \{  H^{(1)}[\delta N_1] \; , \;  H^{(0)}[\bar{N}]  \} \nn \\
& +  \{  H^{(1)}[ \delta N_1] \; , \;  H^{(2)}[\bar{N}] \}  +\{  H^{(2)}[\bar{N}] \; , \;  H^{(1)}[\delta N_2] \}
\end{align}

We first compute the Poisson bracket between $H^{(0)}[\bar{N}]$ and $H^{(1)}[\delta N_1]$. It reads
\bea
& &\{  H^{(1)}[ \delta N_1] \; , \;  H^{(0)}[\bar{N}] \}\nn \\
& & = \frac{1}{4 \kappa^2} \int dx^3 \; \big{\{} \; \delta N_1 ( 2 \sqrt{p} \epsilon^{ab}{}_k \partial_a \delta A^k_b + 4 \sqrt{p} f_1(c) \delta A^i_a\delta^a_i + \frac{1}{\sqrt{p}} f_2(c) \delta E^a_i\delta^i_a) \; ,  6V_0\bar{N} \sqrt{p} g(c) \; \big{\}}\nn \\
& &=  \frac{i}{12 \kappa} \int dx^3 \; \bar{N} \delta N_1 \; \left[ \; -  6 \dot{g}(c) \; \epsilon^{ab}{}_k \partial_a \delta A^k_b + 12 g(c)\dot{f}_1(c) \delta A^i_a\delta^a_i  - 12 \dot{g}(c)f_1(c) \delta A^i_a\delta^a_i\right.\nn\\
& &\hspace{3cm} \left. + 3 p^{-1}\dot{f}_2(c)g(c) \delta E^a_i\delta^i_a + 3 p^{-1}\dot{g}(c) f_2(c) k^3 \delta E^a_i\delta^i_a \; \right]
%& &=  \frac{1}{ \kappa} \int dx^3 \; \bar{N} \delta N_1 \; ( \; -  k \; \epsilon^{ab}{}_k \partial_a \delta A^k_b  - k^2 \delta A^i_a\delta^a_i +  k^3 \delta E^a_i\delta^i_a  \; )
\eea
Taking into account its symmetric contribution coming from $\{  H^{(0)}[\bar{N}] \; , \;  H^{(1)}[\delta N_2] \}$, we have
\bea\label{H0withH1}
& & \{  H^{(0)}[\bar{N}] \; , \;  H^{(1)}[\delta N_2] \}  +  \{  H^{(1)}[\delta N_1] \; , \;  H^{(0)}[\bar{N}]  \} \nn\\
& &=  \frac{i}{4 \kappa} \int dx^3 \; \bar{N} \left[\delta N_2-\delta N_1\right]\; \left[ \; 2 \dot{g}(c) \; \epsilon^{ab}{}_k \partial_a \delta A^k_b -4 g(c)\dot{f}_1(c) \delta A^i_a\delta^a_i  +4 \dot{g}(c)f_1(c) \delta A^i_a\delta^a_i \right.\nn\\
& &\hspace{3cm} \left. +  p^{-1}\dot{f}_2(c)g(c) \delta E^a_i\delta^i_a +  p^{-1}\dot{g}(c) f_2(c) k^3 \delta E^a_i\delta^i_a \; \right]
\eea

Finally, we would like to compute (in the following, we suppress the dependence on the spatial coordinate after the first line, for the sake of brevity)
\bea
& &\{  H^{(1)}[ \delta N_1] \; , \;  H^{(2)}[\bar{N}] \} \nn\\
& &= \frac{1}{4\kappa^2}\int dx^3 dy^3 \delta N_1(x) \bar{N}\bigg\{ \;  \left[2 \sqrt{p} \epsilon^{ab}{}_k \partial_a \delta A^k_b(x) + 4 \sqrt{p} f_1(c) \delta A^m_c(x)\delta^c_m + \frac{f_2(c)}{\sqrt{p}} \delta E^c_m(x)\delta^m_c\right]\, , \nn\\
& &\hspace{1cm} \left[\frac{1}{2} p^{-3/2} h_1 (c) \; \delta E^a_i(y)\delta^i_b \delta E^b_j(y) \delta^j_a - \frac{1}{4} p^{-3/2} h_2(c) (\delta^i_a\delta E^a_i(y) )^2  + \sqrt{p} \;h_4(c) \; (\delta A^i_a(y)\delta^a_i)^2 - \sqrt{p}\; h_4(c)\; \delta A^i_a(y)\delta^a_j \delta A^j_b(y) \delta^b_i \right.\nn  \\
&  &\; \left.+ \frac{2}{\sqrt{p}} \; h_3(c) \; \delta E^a_i(y) \delta A^i_a(y) - \frac{1}{\sqrt{p}}  \; (\delta E^a_i(y)\delta^i_a) \;  \epsilon^{ab}{}_k \; \partial_a \delta A^k_b(y) + \frac{2}{\sqrt{p}} \; \delta^a_i \delta E^b_j(y) \; \epsilon^{ij}{}_k \; \partial_{[a} \delta A^k_{b]}(y)\right]  \bigg\} \nn\\
&  &= \frac{1}{4\kappa^2}\int dx^3 dy^3 \bar{N}\partial_a \delta N_1 \bigg\{ -   \epsilon^{ab}{}_k \delta A^k_b \; , \left[  \frac{h_1(c)}{p} \delta E^c_m \delta^m_d \delta E^d_n \delta^n_c - \frac{h_2(c)}{2p} (\delta E^c_m \delta^m_c )^2 + 4 h_3(c) \delta E^c_m \delta A^m_c \right.\nn  \\
&  &\qquad \qquad \qquad \qquad \qquad \qquad \qquad \qquad \;\;\; \left.- 2 \delta^m_c \delta E^c_m \epsilon^{ef}{}_n \partial_e \delta A^n_f  + 4 \delta^e_i \delta E^f_j \epsilon^{ij}{}_n \partial_{[e} \delta A^n_{f]}  \; \right]  \bigg\}_{(\delta A, \delta E)} \nn\\
& &+ \frac{1}{4\kappa^2}\int dx^3 dy^3  \bar{N}\delta N_1 f_1(c)\bigg\{ \;    \delta^g_p \delta A^p_g \; , \; \left[\frac{2h_1(c)}{p} \delta E^c_m \delta^m_d \delta E^d_n \delta^n_c  - \frac{h_2(c)}{p}(\delta E^c_m \delta^m_c )^2 + 8 h_3(c) \delta E^c_m \delta A^m_c\right. \nn\\
& &\qquad \qquad \qquad \qquad \qquad \qquad \qquad \qquad \;\;\; \left.-4   \delta^q_u \delta E^u_q \epsilon^{ab}{}_k \partial_a \delta A^k_b + 8  \delta^a_i \delta E^b_j \epsilon^{ij}{}_k \partial_{[a} \delta A^k_{b]} \;\right] \bigg\}_{(\delta A, \delta E)}\nn \\
& &+ \frac{1}{4\kappa^2}\int dx^3 dy^3  \bar{N} \delta N_1 f_2(c) \bigg\{ \; \delta^p_g \delta E^g_p \; , \;\left[  \; h_4(c)(\delta A^m_c \delta^c_m)^2 - h_4(c)\delta A^m_c \delta^c_n \delta A^n_d \delta^d_m + \frac{2 h_3(c)}{p} \delta E^c_m\delta A^m_c \right.\nn \\
& &\qquad \qquad \qquad \qquad \qquad \qquad \qquad \qquad \;\;\;\left. - \frac{1}{p} \delta E^c_m\delta^m_c \;  \epsilon^{ab}{}_k \partial_a \delta A^k_b +  \frac{2}{p}  \delta^a_i \delta E^b_j \epsilon^{ij}{}_k \partial_{[a} \delta A^k_{b]} \;\right] \bigg\}_{(\delta A, \delta E)}\nn \\
& &+ \frac{1}{4\kappa^2} \int dx^3 dy^3 \{ \cdots , \cdots \}_{(c,p)}
\eea
In the above calculations the subscript of the Poisson brackets suggest the variables with which they are taken, background or perturbative. The very last term, where the Poisson bracket is taken with respect to the background variables, turns out to be of higher than second order so we disregard it in our calculation. From now onwards, we also suppress the argument of the holonomy correction functions. Proceeding with the evaluation of the Poisson brackets
\bea
& &\{  H^{(1)}[ \delta N_1] \; , \;  H^{(2)}[\bar{N}] \} \nn\\
& &= \frac{i}{4\kappa} \int dx^3 \; \left(-\bar{N} \partial_a \delta N_1\right) \; \left( \; \frac{h_1}{p} \epsilon^{ab}{}_k \delta^c_b \delta^k_m \delta^n_c \delta E^d_n \delta^m_d +  \frac{h_1}{p} \epsilon^{ab}{}_k \delta^d_b \delta^k_n \delta^m_d \delta E^c_m \delta^n_c - \frac{h_2}{2p} \; 2 \epsilon^{ab}{}_k \delta^c_b \delta^k_m \delta^m_c \delta A^i_d\delta^d_i \right.\nn \\
& &\left. + 4 h_3 \epsilon^{ab}{}_k \delta^k_m \delta^c_b \delta A^m_c - 2 \epsilon^{ab}{}_k \delta^m_c \delta^k_m \delta^c_b \epsilon^{ef}{}_n \partial_e \delta A^n_f +  4 \epsilon^{ab}{}_k \delta^e_i \delta^k_j \delta^f_b \epsilon^{ij}{}_n \partial_{[e} \delta A^n_{f]} \; \right) \nn\\
& &+ \frac{i}{4\kappa} \int dx^3 \; \left(\bar{N} \delta N_1\right) \; \left( \; \frac{2f_1 h_1}{p} \delta^g_p \delta^m_g \delta^p_c \delta^c_n \delta E^n_d \delta^d_m + \frac{2f_1 h_1}{p} \delta^g_p \delta E^m_c \delta^c_n \delta^n_g \delta^p_d \delta^d_m - \frac{2f_1 h_2}{p} \delta^g_p \delta^c_g \delta^p_m \delta^m_c \delta E \right. \nn \\
& &\left.+ 8 f_1h_3 \delta^g_p \delta^p_c \delta^m_g \delta^c_n \delta A^n_d \delta^d_m - 4 f_1 \delta^g_p \delta^q_u \delta^u_g \delta^p_q \epsilon^{ab}{}_k \partial_{a} \delta A^k_b + 8 f_1 \delta^g_p \delta^a_i \delta^b_g \delta^p_j \epsilon^{ij}{}_k \partial_{[a} \delta A^k_{b]} - 2 f_2 h_4 \delta^c_m \delta^m_g \delta^p_c \delta^g_p \delta A \right. \nn\\
& &\left.+ f_2 h_4 \delta^g_p \delta^m_g \delta^p_c \delta^c_n \delta A^n_d \delta^d_m + f_2 h_4 \delta^g_p \delta A^m_c \delta^c_n \delta^n_g \delta^p_d \delta^d_m - \frac{2f_2h_3}{p} \delta^g_p \delta E^m_c \delta^c_n \delta^n_g \delta^p_d \delta^d_m \right. \nn\\
& &\left.  -  \frac{f_2}{p} \delta^g_p \epsilon^{ab}{}_k \delta^p_b \delta^k_g \partial_a \delta E^c_m\delta^m_c + \frac{2f_2}{p} \delta^g_p \delta^a_i \delta^k_g \delta^p_b \epsilon^{ij}{}_k \partial_a \delta E^b_j + \frac{2f_2}{p} \delta^g_p \delta^a_i \delta^k_g \delta^p_a \epsilon^{ij}{}_k \partial_b \delta E^b_j \; \right)
\eea

Our aim is to separate the terms proportional to the derivative of the (perturbed) shift vector, $(\bar{N} \partial_a\delta N_1)$, and the rest of the terms, which are proportional to $(\bar{N}\delta N_1)$. The reason for this shall be clear later.
\bea
& &\{  H^{(1)}[ \delta N_1] \; , \;  H^{(2)}[\bar{N}] \} \nn\\
& &= \frac{i}{4\kappa} \int dx^3 \; (- \bar{N} \partial_a \delta N_1) \; \left( \; \frac{2h_1}{p} \epsilon^{ab}{}_k \delta E^k_b + 4h_3 \epsilon^{ab}{}_k \delta A^k_a + 4 \epsilon^{ab}{}_j \epsilon^{ej}{}_n \partial_{[e} \delta A^n_{b]} \; \right)\nn \\
& &+ \frac{i}{4\kappa}  \int dx^3 \bar{N} \delta N_1 \; \left( \; - \frac{1}{p} [4f_1h_1 - 6f_1h_2 - 2f_2h_3] \delta E^a_i\delta^i_a + [8f_1h_3 - 4 f_2h_4] \delta A^i_a\delta^a_i\right. \nn\\
& &\hspace{4cm} \left. + \,4 f_1 \epsilon^{ab}{}_k \partial_a \delta A^k_b - \frac{2f_2}{p} \epsilon_{ab}{}^{k} \partial^a \delta E^b_k \; \right)\nn \\
& &= \frac{i}{4 \kappa} \int dx^3 \; \left(- \bar{N} \partial_a \delta N_1\right) \; \left( \; \frac{2}{p}[h_1 + f_2] \epsilon^{a}{}_{b}{}^k \delta E^b_k + 4 h_3 \epsilon^{ab}{}_k \delta A^k_a + 4 \partial^a \delta A^i_c\delta^c_i  - 4 \delta^i_b \partial^b  \delta A^a_i\; \right) \nn\\
& &+ \frac{i}{4\kappa}  \int dx^3 \bar{N} \delta N_1 \; \left( \; - \frac{1}{p} [4f_1h_1 - 6f_1h_2 - 2f_2h_3] \delta E^a_i\delta^i_a + [8f_1h_3 - 4 f_2h_4] \delta A^i_a\delta^a_i\right.\nn\\
& & \hspace{4cm} \left.  +\, 4 f_1 \epsilon^{ab}{}_k \partial_a \delta A^k_b  \; \right)\nn\\
& &=  \frac{i}{ \kappa} \int dx^3 \; \left(- \bar{N} \partial_a \delta N_1\right) \; \left(\frac{1}{p}\right) \; \left( \; \frac{[h_1 + f_2]}{2}\, \epsilon^{a}{}_{b}{}^k \delta E^b_k + p\, h_3\, \epsilon^{ab}{}_k \delta A^k_a +  p\,\partial^a \delta A^i_c\delta^c_i   - p\, \delta^i_b \partial^b  \delta A^a_i\; \right)\nn\\
& & + \frac{i}{4\kappa}  \int dx^3 \bar{N} \delta N_1 \; \left( \; - \frac{1}{p} [4f_1h_1 - 6f_1h_2 - 2f_2h_3] \delta E^a_i\delta^i_a + [8f_1h_3 - 4 f_2h_4] \delta A^i_a\delta^a_i \right.\nn\\
& &\hspace{4cm} \left. +\, 4 f_1 \epsilon^{ab}{}_k \partial_a \delta A^k_b  \; \right)
%& &= D^{(1)}[- \bar{N} \partial_a \delta N_1] + \frac{1}{\kappa}  \int dx^3 \bar{N} \delta N_1 \; ( \; - \frac{k^3}{p} \delta E +  k^2 \delta A +  k \epsilon^{ab}{}_k \partial_a \delta A^k_b  \; )\nn
\eea
Once we include the contribution coming from $\{  H^{(2)}[\bar{N}\;] , \;  H^{(1)}[ \delta N_2]  \}$, we have the full bracket
\bea\label{H2withH1}
& & \{  H^{(2)}[\bar{N}] \; , \;  H^{(1)}[\delta N_2]  \} + \{  H^{(1)}[\delta N_1] \; , \;  H^{(2)}[\bar{N}]  \}\nn\\
& & = \frac{i}{ \kappa} \int dx^3 \; \bar{N} \left(\partial_a \delta N_2 -  \partial_a \delta N_1\right) \; \left(\frac{1}{p}\right) \; \left( \; \frac{[h_1 + f_2]}{2}\, \epsilon^{a}{}_{b}{}^k \delta E^b_k + p\, h_3\, \epsilon^{ab}{}_k \delta A^k_a +  p\,\partial^a \delta A^i_c\delta^c_i   - p\, \delta^i_b \partial^b  \delta A^a_i\; \right)\nn\\
& & + \frac{i}{4\kappa}  \int dx^3 \bar{N} \left(\delta N_1 - \delta N_2\right) \; \left( \; - \frac{1}{p} [4f_1h_1 - 6f_1h_2 - 2f_2h_3] \delta E^a_i\delta^i_a + [8f_1h_3 - 4 f_2h_4] \delta A^i_a\delta^a_i \right.\nn\\
& &\hspace{4cm} \left. +\, 4 f_1 \epsilon^{ab}{}_k \partial_a \delta A^k_b  \; \right)\,.
\eea

We first consider the terms proportional to $\bar{N} \left(\delta N_1 - \delta N_2\right)$ coming from (\ref{H2withH1}), and similar terms coming from (\ref{H0withH1}). In the classical calculation, these terms cancel each other and following that logic, we impose conditions on the holonomy correction functions
\bea
& & -\frac{1}{p} [4f_1h_1 - 6f_1h_2 - f_2h_3] \delta E^a_i\delta^i_a + [8f_1h_3 - 4 f_2h_4] \delta A^i_a\delta^a_i + 4 f_1 \epsilon^{ab}{}_k \partial_a \delta A^k_b \nn\\
&=& 2 \dot{g}(c) \; \epsilon^{ab}{}_k \partial_a \delta A^k_b -4 g(c)\dot{f}_1(c) \delta A^i_a\delta^a_i  +4 \dot{g}(c)f_1(c) \delta A^i_a\delta^a_i +  p^{-1}\dot{f}_2(c)g(c) \delta E^a_i\delta^i_a\nn\\
& & +  p^{-1}\dot{g}(c) f_2(c) k^3 \delta E^a_i\delta^i_a\,.
\eea
Cancelling similar terms, we get the conditions
\bea\label{cond1}
2f_1h_3 - f_2h_4 &=& \dot{g}f_1 - g\dot{f}_1 2 \\
f_1 &=& \dot{g} \\
6f_1h_2 + 2f_2h_3 - 4f_1h_1 &=& g\dot{f}_2 + \dot{g}f_2
\eea
It is easy to check that these conditions are obviously satisfied by the classical limits of the correction functions. The terms we are left with from (\ref{H2withH1}) should give us the diffeomorphism constraint (as it did in the classical case). This tells us that the diffeomorphism constraint must also be modified by similar holonomy correction functions.

Therefore, putting all the terms together, we obtain:
\bea
\{ H[N_1], H[N_2]\}  & = & \{  H^{(0)}[\bar{N}] \; , \;  H^{(1)}[\delta N_2] \}  +  \{  H^{(1)}[\delta N_1] \; , \;  H^{(0)}[\bar{N}]  \}\nn \\
& &+  \{  H^{(1)}[ \delta N_1] \; , \;  H^{(2)}[\bar{N}] \}  +\{  H^{(2)}[\bar{N}] \; , \;  H^{(1)}[\delta N_2] \}\nn \\
\Rightarrow \{ H_{\text{grav}}[N_1], H_{\text{grav}}[N_2]\} &=& D_{\text{grav}}\left[\frac{1}{p} \bar{N}\partial_a\left(\delta N_2 - \delta N_1 \right) \right]\,,
\eea
where the modified diffeomorphism constraint is defined in (\ref{modDiffeo}) with the correction functions in the diffeomorphism constraint being related to the ones in the Hamiltonian constraint via the relations
\bea\label{cond2}
\tilde{f}_1 &=& h_3\,, \\
\tilde{f}_2 &=& \frac{[h_1 + f_2]}{2}\,.
\eea

This result is remarkable since this shows that the modified gravitational constraints end up with the same \textit{undeformed} hypersurface deformation algebra as in the classical case, when using self-dual Ashtekar variables. This is quite the opposite to the case of the real variables as has been discussed before.

\subsection{The $\{ H_{\text{grav}}, D_{\text{grav}}\} $ bracket}

Since we end up modifying the diffeomorphism constraint, it is also important to check the other brackets of the full hypersurface deformation algebra to make sure that there are no anomalies. It is easy to convince oneself that this bracket reduces to
\bea
\left\{ H[N]\;, \; D[N^a] \right\} = \frac{1}{2i\kappa^2}\int d^3 x d^3 y \; \delta N(x) \;\delta N^a(y) \left\{  \mathcal{H}^{(1)} \; , \; \mathcal{D}^{(1)} \right\}\,.
\eea
When we evaluate this bracket, we get
\bea\label{HwithD}
& & \left\{ H[N]\;, \; D[N^a] \right\}\nn\\
& &= \frac{1}{2i\kappa^2}\int d^3 x d^3 y \;\delta N(x) \; \delta N^c(y) \left(\left\{ p^{-1/2} f_2 \delta E^a_i\delta^i_a \; ,\; p \partial_c \delta A^j_d\partial^d_j \right\}\right.\nn\\
& &\left.\hspace{1cm} - \left\{ p^{-1/2} f_2 \delta E^a_i\delta^i_a \; ,\; p \delta^d_i\partial_d \delta A^j_c \right\} + 2\left\{ p^{1/2} \epsilon^{ab}{}_k\partial_a \delta A^k_b \; ,\; \tilde{f}_2 \epsilon^i{}_{cd} \delta E^d_i\right\}\right)\,.
\eea
Evaluating the Poisson brackets and after a couple of integration by parts, we get
\bea
 \left\{ H[N]\;, \; D[N^a] \right\}
& =& -\frac{1}{2\kappa} \left[\int d^3 x \, \delta N^a \partial_a \delta N\right]\; \sqrt{p}\;\left(4\tilde{f}_2 + 2f_2\right)\nn\\
\Rightarrow \left\{ H_{\text{grav}}[N]\;, \; D_{\text{grav}}[N^a] \right\} &=&-H_{\text{grav}}\left[\delta N^a \partial_a \delta N\right]\,.
\eea
Classically, we must get the zeroth order Hamiltonian apart from the smearing function, which gives us the condition
\bea\label{cond3}
4\tilde{f}_2 + 2f_2 &=& 6 g\nn\\
\Rightarrow 2(h_1 + f_2)+ 2f_2 &=& 6 g\nn\\
\Rightarrow  h_1 + 2f_2 &=& 3 g\,,
\eea
where in the second equality we have used relation (\ref{cond2}). Once again, it is easy to check the trivial consistency condition that these relations are satisfied by the classical limit of these functions. Thus, together, we have five consistency conditions, Eqs. (\ref{cond1}), (\ref{cond2}), \& (\ref{cond3}) so far for the eight undetermined holonomy correction functions introduced in Eqs. (\ref{modHam}) \& (\ref{modDiffeo}). However, we shall see that there are going to be more constraints appearing on these functions from the matter sector as well.

\subsection{The $\{ D_{\text{grav}}, D_{\text{grav}}\} $ bracket}

Although the diffeomorphism constraints have been modified via holonomy correction functions of the background connection, the Poisson bracket between two diffeomorphism constraints still remains zero, as in the classical case and does not impose any further conditions on our correction functions.

\section{Anomaly free algebra of the matter effective phase space }

Note that the connection $c$ doesn't appear in the matter constraints involving a minimally coupled scalar and there is therefore no modifications to introduce in the constraints. Only the metric (or equivalently, the triad) appears in the scalar field constraints and hence they should not have any holonomy corrections in them. Thus there are no deformations in the matter part of the constraint algebra
\bea\label{matterAlg}
\left\{D_{\text{scal}}[N_1^a]\,,\,D_{\text{scal}}[N_2^a]\right\} &=& 0\\
\left\{H_{\text{scal}}[N]\,,\,D_{\text{scal}}[N^a]\right\} &=& - H_{\text{scal}}[\delta N^a \partial_a \delta N]\\
\left\{H_{\text{scal}}[N_1]\,,\,H_{\text{scal}}[N_2]\right\} &=& D_{\text{scal}}\left[\frac{\bar{N}}{p} \partial^a \left(\delta N_2 - \delta N_1\right)\right]\,.
\eea

\section{Covariance of cosmological perturbations in the self dual formalism}
In this section, we compute the hypersurface deformation algebra for our symmetry reduced model, which has local degrees of freedom through the cosmological perturbations $\delta A^i_a,$ $\delta E^a_i,$ $\delta\Psi$ and $\delta\pi$.

\subsection{The $\{ H_{\text{tot}}, H_{\text{tot}}\}$ bracket }

The total Hamiltonian constraint can be written as
\bea
H_{\text{tot}}[N] = H_{\text{grav}}[N] + H_{\text{scal}}[N]\,,
\eea
and the total diffeomorphism constraint as
\bea
D_{\text{tot}}[N] = D_{\text{grav}}[N] + D_{\text{scal}}[N]\,.
\eea

The Poisson brackets between two (holonomy corrected) total Hamiltonian constraints can be decomposed as
\bea
\left\{H_{\text{tot}}[N_1]\,,\,H_{\text{tot}}[N_2]\right\} &=& \left\{H_{\text{grav}}[N_1]\,,\,H_{\text{grav}}[N_2]\right\} + \left\{H_{\text{scalar}}[N_1]\,,\,H_{\text{scalar}}[N_2]\right\}\nn\\
& &\;\; \left(\left\{H_{\text{grav}}[N_1]\,,\,H_{\text{scalar}}[N_2]\right\} - (N_1 \leftrightarrow N_2)\right)\,.
\eea
While the first two terms has been calculated in the previous sections, and gives us the diffeomorphism constraint with the (undeformed) classical structure function, the last term in the second line of the above equation needs to be evaluated. For the classical calculation, this Poisson bracket goes to zero. Indeed, as we shall see below, requiring that this cross-term goes to zero shall impose further constraints on our holonomy correction functions.

We begin by decomposing the cross-term between the gravitational part of the Hamiltonian part and the scalar part as follows
\bea
& &\left\{H_{\text{grav}}[N_1]\,,\,H_{\text{scalar}}[N_2]\right\} +\left\{H_{\text{scalar}}[N_1]\,,\,H_{\text{grav}}[N_2]\right\}\nn\\
& &= \left\{H^{(1)}_{\text{grav}}[\delta N_1]\, ,\,H^{(0)}_{\text{scalar}}[\bar{N}]\right\}_{(c,p)} + \left\{H^{(0)}_{\text{grav}}[\bar{N}]\, , \, H^{(1)}_{\text{grav}}[\delta N_2]\right\}_{(c,p)}\nn\\
& &+ \left\{H^{(1)}_{\text{grav}}[\delta N_1]\, , \, H^{(2)}_{\text{scal}}[\bar{N}]\right\}_{(\delta A,\delta E)} + \left\{H^{(2)}_{\text{grav}}[\bar{N}]\, , \,H^{(1)}_{\text{scal}}[\delta N_2] \right\}_{(\delta A,\delta E)} + (\delta N_1 \leftrightarrow \delta N_2)\,.
\eea

We calculate each of these brackets individually below. Since these calculations are of the exact same nature as those shown above, we only quote the result without getting into details of the algebra. The first bracket is
\bea\label{A1}
& &\left\{H^{(1)}_{\text{grav}}[\delta N_1]\, ,\,H^{(0)}_{\text{scalar}}[\bar{N}]\right\}_{(c,p)} + (\delta N_1 \leftrightarrow \delta N_2)\nn\\
&=& i\int d^3 x\, \bar{N}(\delta N_1 - \delta N_2) \left[p \dot{f}_1 V\;\delta A^i_a\delta^a_i - \frac{1}{2} \dot{f}_1 \left(\frac{\pi^2}{p^2}\right)\delta A^i_a\delta^a_i\right.\nn\\
& &\hspace{3cm} \left. + \frac{1}{4} \dot{f}_2 V\,\delta E^a_i\delta^i_a - \frac{1}{8} \dot{f}_2 \left(\frac{\pi^2}{p^2}\right) \delta E^a_i\delta^i_a \right].
\eea
The next bracket reads
\bea\label{A3}
& &\left\{H^{(0)}_{\text{grav}}[\bar{N}]\, , \, H^{(1)}_{\text{grav}}[\delta N_2]\right\}_{(c,p)}+ (\delta N_1 \leftrightarrow \delta N_2)\nn\\
&=& i\int d^3 x\, \bar{N}(\delta N_1 - \delta N_2) \left[-\frac{3}{2} \dot{g} \left(\frac{\pi\delta \pi}{p^2}\right) + \frac{5}{8}\dot{g}\left(\frac{\pi^2}{p^3}\right)\delta E^a_i\delta^i_a + \frac{3}{2}\dot{g}\, p\, V_{,\Psi} \delta\Psi + \frac{1}{4} \dot{g}\, V\, \delta E^a_i\delta^i_a\right].
\eea
For the next two Poisson brackets, we only write down those terms which are modified due to the presence of holonomy corrections. Specifically, the terms which are proportional to $\delta_a(\delta N_2 - \delta N_2)$ are not explicitly written down since they turn out to be the same as that in the classical case. So these terms get cancelled with each other just the same as in the classical case. Proceeding with the next Poisson bracket leads to
\bea\label{A2}
& &\left\{H^{(1)}_{\text{grav}}[\delta N_1]\, , \, H^{(2)}_{\text{scal}}[\bar{N}]\right\}_{(\delta A,\delta E)}+ (\delta N_1 \leftrightarrow \delta N_2)\nn\\
&=& i\int d^3 x\, \bar{N}(\delta N_1 - \delta N_2) \left[-3f_1\left(\frac{\pi\delta\pi}{p^2}\right) + \left(\frac{3}{4} + \frac{1}{2}\right) f_1 \left(\frac{\pi^2}{p^3}\right) \delta E^a_i\delta^i_a + 3 f_1\; p\;V_{,\Psi}\;\delta\Psi\right.\nn\\
& &\hspace{4cm}\left. \left(\frac{3}{2} - 1\right) f_1\, V\,\delta E^a_i\delta^i_a \right]\,.
\eea
Finally, the last bracket is given by
\bea\label{A4}
& &\left\{H^{(2)}_{\text{grav}}[\bar{N}]\, , \,H^{(1)}_{\text{scal}}[\delta N_2] \right\}_{(\delta A,\delta E)} + (\delta N_1 \leftrightarrow \delta N_2)\nn\\
&=& i\int d^3 x\, \bar{N}(\delta N_1 - \delta N_2) \left[\left(-\frac{3}{2} + \frac{1}{2}\right)h_4\; p\;V\;\delta A^i_a\delta^a_i + \left(\frac{3}{4} - \frac{1}{4}\right)h_4\left(\frac{\pi^2}{p^2}\right)\delta A^i_a\delta^a_i\right.\nn\\
& &\hspace{4cm}\left. +\frac{1}{4} h_3\left(\frac{\pi^2}{p^3}\right) \delta E^a_i\delta^i_a - \frac{1}{2} h_3\;V\; \delta E^a_i\delta^i_a \right]
\eea

On requiring that the cross-term vanishes, just as in the classical case, we get six new conditions on the holonomy correction functions. However, not all of them are independent. For instance, we recover the relation $2 f_1 = \dot{g}$, which we have from before (\ref{cond1}) and can be seen as a nice consistency check. We are left with the following independent new conditions (some of which also appear multiple times)
\bea\label{cond4}
\dot{f}_2 &=& 2 h_3\\
\dot{f}_1 &=& h_4\,.
\eea

\subsection{The $\{ H_{\text{tot}}, D_{\text{tot}}\}$ bracket }

The Poisson bracket between the total Hamiltonian and total diffeomorphism constraints can be decomposed as
\bea\label{fullHwithD}
\left\{H_{\text{tot}}\;,\; D_{\text{tot}}\right\} &=& \left\{H_{\text{scal}}\;,\; D_{\text{scal}}\right\} + \left\{H_{\text{grav}}\;,\; D_{\text{grav}}\right\}\nn\\
& &\;\; \left\{H_{\text{grav}}\;,\; D_{\text{scal}}\right\} + \left\{H_{\text{scal}}\;,\; D_{\text{grav}}\right\}\,.
\eea
The two Poisson brackets involved in the first line, namely the matter part of the Hamiltonian with the matter part of the diffeomorphism constraint and the gravitational part of the Hamiltonian with the gravitational part of the diffeomorphism constraint, have already been calculated in the preceding sections. The terms in the second line of the Eq. (\ref{fullHwithD}) above are cross-terms between gravitational and matter parts of Hamiltonian and diffeomorphism constraints. It is a straightforward, albeit arduous, calculation to show that these two terms indeed vanish just as in the classical case, without giving us any more constraints on the holonomy correction functions.

\subsection{The $\{ D_{\text{tot}}, D_{\text{tot}}\}$ bracket }

The Poisson bracket between the full diffeomorphism constraint with itself is trivially zero as in the classical case.

\subsection{Summary of the precedent computations}

The final form of the full hypersurface deformation algebra takes the form
\bea\label{fullAlgebra}
\left\{D_{\text{tot}}[N_1^a]\,,\,D_{\text{tot}}[N_2^a]\right\} &=& 0\\
\left\{H_{\text{tot}}[N]\,,\,D_{\text{tot}}[N^a]\right\} &=& - H_{\text{tot}}\,[\delta N^a \partial_a \delta N]\\
\left\{H_{\text{tot}}[N_1]\,,\,H_{\text{tot}}[N_2]\right\} &=& D_{\text{tot}}\left[\frac{\bar{N}}{p} \partial^a \left(\delta N_2 - \delta N_1\right)\right]\,.
\eea
where both the Hamitonian and diffeomorphism constraints for the gravtational sector has been modified by holonomy correction functions whereas those in the matter sector remains the same as in the classical case. In addition, we have several consistency condition which have been imposed on the various arbitrary correction functions via relations, Eqs. (\ref{cond1}), (\ref{cond2}), (\ref{cond3}), \& (\ref{cond4}). However, crucially, the full constraint algebra remains undeformed, i.e. they appear with the same structure functions as in the classical case, unlike in the case of using real Ashtekar-Barbero variables. One of the main technical reasons for this happening is that the terms proportional to the spatial derivative of the (perturbed) spin connection, which appear in the real formalism, are absent in the self dual one from the expressions of the Hamiltonian and diffeomorphism constraints. Such terms have been shown to be at the heart of structure function deformation (and consequently, \textit{signature change}) in models of loop quantum gravity \cite{Bojowald:2014zla}. Thus we do not have these features in the self dual formalism of cosmological perturbations, as expected from related work on spherical symmetry \cite{SDSS}.

\subsection{Non trivial holonomy-corrected solutions}\label{Everythinging}

Although we have shown that we can introduce holonomy modification functions in both the Hamiltonian and diffeomorphism constraints, and have a consistent algebra of constraints which remains closed, it might still turn out that the only allowed solution is the classical one. Since we only introduce point-wise holonomy corrections for the background connection alone, and do not polymerize the inhomogeneous perturbations, it is entirely possible that these local correction functions turn out to be the trivial ones from GR, once the restrictions derived in Eqs. (\ref{cond1}), (\ref{cond2}), (\ref{cond3}), \& (\ref{cond4}) are imposed. We shall show in this section that this is not the case and we do have nontrivial holonomy corrections which can be consistently implemented.

Starting with (\ref{cond1}), we can eliminate one of the conditions and  rewrite them as
\bea\label{cond11}
f_1 &=& \frac{1}{2} \dot{g} = f_1\left(\dot{g}\right)\,\\
\dot{g} h_3 - f_2 h_4 &=& \frac{1}{2} \dot{g}^2 - \frac{1}{2} g \ddot{g}\,\\
3\dot{g} h_2 + 2 f_2 h_3 - 2 \dot{g} h_1 &=& g \dot{f}_2 + \dot{g} f_2\,.
\eea
Using (\ref{cond4}) and (\ref{cond3}), we can further simplify them as
\bea\label{cond12}
\dot{g} \left(\dot{f}_2 - \dot{g}\right) + \ddot{g} \left(g - f_2\right) &=& 0\nn\\
\Rightarrow f_2 &=& f_2\left(g, \dot{g}, \ddot{g}\right)\,\label{eq:f2} \\
3\dot{g} \left(h_2 + f_2 - 2g\right) + \dot{f}_2 \left(f_2 - g\right) &=& 0\nn\\
\Rightarrow h_2 = \tilde{h}_2\left(g, f_2, \dot{f}_2, \dot{g}\right) &=& h_2\left(g, \dot{g}, \ddot{g}\right)\,.
\eea
Now, using these equations (\ref{cond12}), we can reformulate (\ref{cond4}) and (\ref{cond3}) as
\bea\label{cond13}
h_3 =\frac{1}{2} \dot{f}_2 &=& h_3 \left(g, \dot{g}, \ddot{g}\right)\,\\
h_4 = \dot{f}_1 &=& h_4 \left(\ddot{g}\right)\,,\\
h_1 = 3g - 2f_2 &=& h_1 \left(g, \dot{g}, \ddot{g}\right)\,.
\eea
Finally, it is obvious that the holonomy modification functions appearing in the diffeomorphism constraint (\ref{cond2}) can easily be recast as
\bea\label{cond14}
\tilde{f}_1 &=& \tilde{f}_1 \left(g, \dot{g}, \ddot{g}\right)\,\\
\tilde{f}_2 &=& \tilde{f}_2 \left(g, \dot{g}, \ddot{g}\right)\,.
\eea

This set of equations can be solved to express all the quantum modifications as functions of $g$ and its derivatives. First one notes that the general solution for Eq. (\ref{eq:f2}) is
\bea
f_2=\lambda\dot{g}+g,
\eea
with $\lambda$ an arbitrary integration constant. Using this as well as $f_1=\dot{g}/2$, it is straighforward to extract all the other modification functions. For the modifications entering the Hamiltonian constraint, this gives
\bea
h_1&=&g-2\lambda\dot{g}\, , \\
h_2&=&g-\frac{2\lambda}{3}\dot{g}-\frac{\lambda^2}{3}\ddot{g}\, , \\
h_3&=&\frac{1}{2}\left(\lambda\ddot{g}+\dot{g}\right)\, , \\
h_4&=&\frac{1}{2}\ddot{g}\, .
\eea
The modification functions entering the diffeomorphism constraints are obtained from Eq. (\ref{cond2}), yielding
\bea
\tilde{f}_1&=&\frac{1}{2}\left(\lambda\ddot{g}+\dot{g}\right)\, , \\
\tilde{f}_2&=&g-\frac{\lambda}{2}\dot{g}\, .
\eea
Finally, the value of the integration constant $\lambda$ can be set by requiring that GR is recovered in the classical limit. This imposes that $g(c)\to c^2$, $f_1(c)\to c$, and $f_2(c)\to c^2$, for $c\to0$. Considering the $g$ function to be analytic, we can write it as an expansion in the classical regim, $g(c)\sim c^2+\mu c^3+\mathcal{O}(c^4)$. The proper classical limit of $f_1$ is straighforwardly obtained from $f_1=\dot{g}/2$ and the classical expansion of $g$, i.e. $f_1\sim c+ (3\mu/2) c^2+\mathcal{O}(c^3)$. For the second function, $f_2$, this gives $f_2\sim 2\lambda c +c^2+3\mu\lambda c^2+\mathcal{O}(c^3)$. The classical limit, $f_2\sim c^2$, is thus obtained by setting the integration constant $\lambda=0$. In fact, obtaining coherently the classical limit for $g$, $f_1$, and $f_2=2\lambda f_1+g$, imposes to set $\lambda=0$.

This shows that  all the correction functions can be written in terms of the holonomy modification function appearing in the homogeneous (self-dual) LQC equation, $g(c)$. In the case of the real variables, this would mean that all the other functions can be written in terms of the $\sin\left(\delta c\right)/\delta$ function arising from the holonomy regularization in LQC. Similarly, once we know the corresponding form for the self-dual Ashtekar variables, we can easily recast the rest of the functions in terms of it. Interestingly, they admit a rather simple form as functions of $g$ and its derivatives:
\bea
f_1=h_3=\tilde{f}_1&=&\frac{1}{2}\dot{g}\, , \\
f_2=h_1=h_2=\tilde{f}_2&=&g\, ,\\
h_4&=&\frac{1}{2}\ddot{g}\, .
\eea
More importantly, this serves our initial purpose of proving that we include genuine quantum corrections since our form of the constraints are different from those one gets in classical GR. This is obvious from the fact that we can choose any arbitrary form of the function $g(c)$ (and not necessarily the classical limit of $c^2$) and our results would still be valid. Additionally, this shows that the classical limit is automatically satisfied for our consistency conditions.

%In the above manipulations, we do not explicitly solve these differential equations but show that all the correction functions can be formally written in terms of the holonomy modification function appearing in the homogeneous (self-dual) LQC equation, $g(c)$. In the case of the real variables, this would mean that all the other functions can be written in terms of the $\sin\left(\delta c\right)/\delta$ function arising from the holonomy regularization in LQC. Similarly, once we know the corresponding form for the self-dual Ashtekar variables, we can easily recast the rest of the functions in terms of it. More importantly, this serves our initial purpose of proving that we include genuine quantum corrections since our form of the constraints are different from those one gets in classical GR. This is obvious from the fact that we can choose any arbitrary form of the function $g(c)$ (and not necessarily the classical limit of $c^2$) and our results would still be valid. Additionally, this automatically shows that the classical limit is automatically satisfied for our consistency conditions.

\section{Reality conditions}
Since we are using self dual variables, one needs to impose the so called reality conditions in order to recover GR. In the full theory, one of those reality conditions ensure that the metric is indeed real-valued and the second one arises by looking at the stability under the flow of the first constraint, i.e.
\begin{align*}
& g_{ab} = E^i_a E_{i b}  \in \mathbb{R} \qquad \rightarrow \qquad E^i_a E_{i b} - \bar{E}^i_a \bar{E}_{i b} = 0 \\
& \qquad \qquad \qquad  \qquad \;\;\;\; \rightarrow \qquad  A^i_a + \bar{A}^i_a = 2 \Gamma^i_a.
\end{align*}
We note that in this section, $\bar{G}$ means the complex conjugate of $G$. Both conditions turn out to be second class constraints on the phase space and involve the complex conjugate of the canonical variables, which do not belong to the phase space. Thus, in principle, one should enlarge the phase space first by including the complex conjugates of the canonical self-dual variables \cite{AlexRC}. (Note that the two reality conditions presented above are not unique and one can write them in a way which does not involve the complex conjugate canonical variables ; see \cite{MoralesRC, ThiemannRC} for different ways to implement those reality conditions.)

In our case of a perturbed homogeneous and isotropic universe, several simplifications occur. A general strategy we can adopt in our symmetry reduced model is to ignore, in a first attempt, the imposition of the reality conditions and proceed to the quantization first. Then, if one is able to have a full knowledge of the Dirac observables, one can simply require that those observables be self adjoint on the physical Hilbert space. Let us discuss this strategy more concretely.

We begin by looking at the background variables first. For these complex variables, $(c,p)$, the reality conditions turns out to be trivial because we are restricting our study to a flat open universe, i.e. the background spin connection vanishes. We end up with
\be
( p^2 - \bar{p}^2 )\delta_{ab} = 0 \qquad c + \bar{c} = 0
\ee
It implies that $p \in \mathbb{R}$, and that $c = - \bar{c}$, meaning that $c$ is purely imaginary, which is indeed the case since $c = i \dot{a}/N$. Moreover, applying our strategy to this case would not be difficult, because the only observable we have at the background level is the volume of the universe (or equivalently, the scale factor). So implementing the reality conditions on the background variables can be overcome by requiring a self adjoint volume operator.

Next, we consider the perturbed variables $(\delta A^i_a, \delta E^b_j)$. Since at this level, the perturbed spin connection $\delta \Gamma^i_a$ is not vanishing, the reality conditions for the perturbed variables are no more trivial. However, one can still follow the same strategy than for the background variables. First, one derive the `gauge invariant' quantities which are directly related to observations and then impose the reality conditions on them. Since those fields are quantized using the Schr\"odinger representation, i.e. {\it \`a la} Fock, it is straightforward to proceed along this line. Doing so, one can simply circumvent imposing the non trivial reality conditions, associated to the perturbed variables  $(\delta A^i_a, \delta E^b_j)$, on the kinematical Hilbert space. In following such a strategy, we would first restrict ourselves to a Gaussian state for the cosmological perturbations. In the effective approach, this means calculating the expectation values of $(\delta A^i_a, \delta E^b_j)$ in a state sharply peaked around a Gaussian distribution. (It can be shown that this state remains sharply peaked if we only consider up to the second order Hamiltonian.) Then the relevant Dirac observables for us would be related to the gauge invariant curvature perturbation, using which we can evaluate the power spectrum. We would require these variables to be self-adjoint on the physical Hilbert space, just as required for the background variables. Finally, we can then allow for higher order correlation functions (such as the bispectrum, trispectrum and so on) in a controlled manner. When incorporating these non-Gaussian observables in our analysis, we would be enlarging the physical Hilbert space at each order while simultaneously requiring that the Dirac observables spanning it becomes self-adjoint. This would obviously imply that the wave function for the perturbations is no longer strictly Gaussian, but rather perturbatively non-Gaussian. The vital reason why such a strategy seems implementable is that the number of Dirac observables, at each stage, remains finite\footnote{Obviously the Dirac observables, related to the gauge invariant quantities, are fields and when we say they are finite, it implies that they form a countably finite basis of the (separable) physical Hilbert space of cosmological perturbations.}. A required assumption for the implementation of this technique remains that the full physical Hilbert space can be factored out between that of the background variables and the perturbations, i.e. $\mathcal{H}_\text{phys} = \mathcal{H}_\text{phys}^\text{background} \otimes \mathcal{H}_\text{phys}^\text{pert}$. Note that this is not an extra assumption since we have already assumed a polymer quantization for background variables while preferring a Schr\"odinger one for the perturbations.

%Moreover, if the perturbed quantum field is in a Gaussian vacuum state, all the Dirac observables one can extract can be deduced from its two points function. Therefore, there is a finite number of Dirac observable associated to this gauge invariant field per point {\Jibril NEED TO BE IMPROVED}.

%However, there can be one other approach that can be taken to implement the reality conditions. Recall that in our procedure, the background and the perturbed variables are not quantized through the same representation. While the background variables are quantized through the polymer representation, one treats the perturbed variables through the usual Fock representation on top of the quantum background.

%In LQG, the difficulty to implement the reality conditions are related to the fact that up to now, no one has found a suitable inner product for the self dual quantum theory because of the unboundedness of the $\SL(2,\mathbb{C})$ Ashtekar-Lewandowski measure which turns out to be ill-defined. One would expect that the reality conditions have to be implemented by the insertion of a projector selecting the well defined quantum state but up to now, such projector remains elusive.

%It turns out that this problem is not present in the usual Schrodinger quantization where the inner product of the quantum theory can be build easily, even in the presence of a non compact group of symmetry. Therefore, implementing the reality conditions should be possible because of the representation chosen for the perturbed variables.

\section{Conclusion}
The first conceptual hurdle we have to overcome is how we can avoid the uniqueness theorem due to Kuka\v{r}, Hojman and Teitelboim which states that starting from the (classical) algebra of hypersurface deformations, one can uniquely obtain the Einstein-Hilbert action when one restricts oneself to a maximum of second order derivatives of the geometrodynamical variables \cite{HKT1, HKT2}. However, in this work, we have established a counter-example to it since we have the classical hypersurface deformation algebra and, yet, find solutions which are different from GR. This can be explained by the fact that we are dealing with a symmetry reduced system in this case and there is considerably more freedom for us than what has been demonstrated for full GR. Another interesting difference for us lies in the fact that the gauge transformations generated by our modified spatial diffeomorphism constraint deforms the notion of spatial hypersurfaces in our formalism. We come back to this point later.

The natural extension for our work would be to find a suitable polymerization function for the $SL(2,\mathbb{C})$ gauge group, which can be obtained by regularizing the curvature function in terms of holonomies with inputs from the full LQG theory. This task is, however, not straightforward as it involves several obstructions related to using a non-compact gauge group. Indeed, \`a priori it is not even clear if such a regularization would lead to singularity resolution in homogeneous LQC, as in the case of the real variables. However, some recent results have shown promising signs that singularity resolution might take place even for the self dual variables. The most rewarding approach so far has been that of an `analytic continuation' procedure sending the Immirzi parameter from a real value to the imaginary one, as well as simultaneously analytically continuing the spin representation governing the dynamics \cite{JBACosmo}. This attempt to define self dual LQC through an analytic continuation was based on some previous results obtained in the context of black holes thermodynamics \cite{MG-AC, JBA-AC1, JBA-AC2, JBA-AC3, GeillerNearRad, JBAthesis} and in $(2+1)$ LQG \cite{BTZAC, JBA1, JBA2}, where such procedure was successfully implemented to derive predictions of the self dual loop quantum theory, such as the Bekenstein-Hawking entropy of (spherically symmetric and axisymmetric) isolated horizons. A crucial point is that, as shown in Sec. \ref{Everythinging}, once we have an expression for the background holonomy correction, it is possible to get all the modification functions appearing in the perturbation Hamiltonian and, subsequently, calculate physically observable quantities like the CMB power spectrum and bispectrum for inflationary models, modified by the LQC quantum corrections.

An important step in this program is the derivation of the gauge-invariant variables (such as e.g. the Mukhanov-Sasaki variables) describing the perturbative degrees of freedom, and their associated equation of motion from which primordial power spectra can be predicted. Concrete computations of these spectra obvisouly require to specify the functions encoding quantum corrections. We note however that general expressions for the gauge-invariant variables (as well as their equation of motion) can be derived as solely parametrized by the polymerization function, $g(c)$ \cite{bbginprep}.

Notwithstanding the fact that the form of the function has been kept arbitrary in this article, we are still able to extract several important conclusions regarding holonomy corrections appearing from LQG.

\subsection{Big bounce or Signature change?}
As mentioned in the introduction, recently, there has been two different research directions within LQC with somewhat conflicting predictions. The strictly minisuperspace quantization of homogeneneous FLRW (or other similar cosmological spacetimes) results in resolving the classical singularity and replacing it with a bounce which connects the expanding branch of the Universe with a previously contracting one. Having quantized the background in LQC, one approach has been to reformulate the resulting quantum spacetime in terms of a \textit{dressed, effective} FLRW metric, having made some gauge choices for the time coordinate, on which perturbations can propagate as independent degrees of freedom. On the other hand, the `deformed algebra' approach within LQC begins by quantizing the full inhomogenous model which includes cosmological perturbations on top of a homogneous background, and then requiring that such a quantization preserves some notion of covariance, as has been the philosophy of this article. The underlying quantum spacetime is deformed and can give rise to the phenomenon of signature change, whereby one effectively  goes from a Lorentzian universe to an Euclidean one. Indeed, this implies that one \textit{cannot} use an `effective' Riemannian geometry to describe such quantum spacetimes, for higher curvature regimes. This seems to be in conflict with the `dressed' metric approach, as summarized above.

%The quantum modifications to the background, achieved via loop quantization, affect the cosmological perturbations through the `dressed' scale factor and conformal time coordinate (explicitly, these quantities are expectation values of well-defined operators on the LQC Hilbert space). Finally, the cosmological perturbations are quantized using usual Fock quantization. Thus we have a paradigm of quantum perturbations on top of a \textit{quantum} FLRW background, resulting in quantum-geometry corrections to the equations of motions of the perturbation modes.

%Here the main quantum effects of LQC are implemented as holonomy and inverse-volume %modifications. The most significant achievement of singularity resolution within LQC %comes from the holonomy corrections where, for real variables, the homogeneous %connection can be replaced by a bounded trigonometric function (specifically, the %sine function). Covariance demands that there are no anomalies in the %holonomy-corrected constraint algebra; although in the case of real variables, %modification of structure functions becomes inevitable.

Our formalism starts out being very similar to the `deformed algebra' approach since our guiding principle is to ensure that we have covariance (or a deformed version of it) for the inhomogeneous, cosmological model. However, the crucial difference for us with both the above mentioned approaches is that we use self dual Ashtekar variables as opposed to their real-valued counterparts. This leads us to the astonishing finding that the full hypersurface deformation algebra remains unmodified and we do not get deformed structure functions, and consequently do \textit{not} have signature change. Thus our findings are not very similar to the ones coming from the `deformed algebra' approach although we share similar strategies. From this point of view, it would seem that our approach ends up finally being closer to the `dressed metric' formulation. Yet there are important differences with that as well. One such indication would be that we require that the (spatial) diffeomorphism constraint gets necessarily modified in order for the closure of the algebra. This is slightly unorthodox from the point of view of LQG, especially while working with the spin network basis, as explained earlier. It might be signalling towards the fact that the underlying quantum background cannot indeed be written with an `effective' metric structure, at least not for the whole parameter space. Another concrete difference would be the direct consequence of using self dual variables. Although it is possible to resolve singularity in the minisuperspace model as indicated by \cite{JBACosmo}, the explicit nature of the bounce would end up being rather different from what one has in the `dressed metric' approach. Therefore, it is difficult to conclude whether our formalism is closer to the `bounce' scenario or the `signature change' one. Rather, we might be finding a confluence between the two, sharing certain properties from both. Specific phenomena like signature change, which are not present in our case, might turn out to be gauge-artifacts of using real-valued variables but, on the other hand, we might indeed be lacking an `effective' (pseudo)Riemannian structure in the deep quantum regime (since our diffeomorphism constraint get modified in the process). These issues remain to be examined in more detail and would be confronted in forthcoming work.

\acknowledgments{We thank Martin Bojowald for several discussions regarding the conceptual ramifications of this work. This work has been partially supported by the Shanghai Municipality, through the grant No. KBH1512299, and by Fudan University, through the grant No. JJH1512105.}

\bibliographystyle{apsrev4-1}
\bibliography{refs}

%\end{thebibliography}

\end{document}